\documentclass[preprint,12pt]{elsarticle}

\usepackage{graphicx}\usepackage{multirow}
\usepackage{amsmath,amssymb,amsfonts}
\usepackage{amsthm}\usepackage{mathrsfs}\usepackage[title]{appendix}\usepackage{xcolor}\usepackage{xcolor}\usepackage{colortbl}
\usepackage{textcomp}\usepackage{manyfoot}\usepackage{algorithm}\usepackage{algorithmicx}\usepackage{algpseudocode}\usepackage{listings}\usepackage{longtable}
\usepackage{pifont}

\usepackage{tabularx}
\usepackage{makecell}

\usepackage{tablefootnote}
\usepackage{textcomp}
\usepackage[table]{xcolor}
\usepackage{minted}
\usepackage{hyperref}
\usepackage{enumitem}           \usepackage[T1]{fontenc} \usepackage{tikz} \usepackage{stfloats}
\usepackage[utf8]{inputenc}
\usepackage{soul}
\usepackage{colortbl}
\definecolor{mygray}{rgb}{0.71,0.71,0.71}
\usepackage{balance}
\usepackage{amsfonts}
\usepackage[caption=false]{subfig} \usepackage{array}

\usepackage{mdframed}

\newcolumntype{P}[1]{>{\centering\arraybackslash}p{#1}}

\newcommand{\secref}[1]{Section~\ref{#1}}

\newcommand{\ie}{\textit{i.e.},\ }
\newcommand{\eg}{\textit{e.g.},\ }
\newcommand{\etal}{\textit{et al.} }
\newcommand{\etc}{{\em etc.}}

\usepackage{booktabs}

\definecolor{francBlue}{RGB}{64,76,87}

\usepackage[most]{tcolorbox}
\tcbuselibrary{raster}

\usepackage[parfill]{parskip}

\tcbset{
  parskip/.style={
    before={\par\pagebreak[0]\parindent=0pt},
    after={\parfillskip=0pt\par},
  },
}
\newtcolorbox{resultbox}[1][]{colback=black!3,
    colframe=black!3,
    notitle,
    sharp corners,
    borderline west={2pt}{0pt}{gray!80!black},
    enhanced,
    breakable,
    boxsep=0pt,
    left=4pt,right=2pt,top=2pt,bottom=2pt,
    }
\newcommand{\rques}[1]{  
\begin{tcolorbox}[enhanced jigsaw,colback=white,left=0pt,right=0pt,top=0pt,bottom=0pt]
\textbf{#1}
\end{tcolorbox}
}

\definecolor{codebg}{rgb}{0.99,0.99,0.99}
\definecolor{hiliteColor}{rgb}{1,0.92549019607,0.6}
\definecolor{tainted}{rgb}{0,1,1}

\newmintinline{python}{fontsize=\normalsize}

\newminted[PythonSourceCode]{python}{
    labelposition=topline,
    frame=single,
    numbersep=2pt,
    fontsize=\tiny,
    xleftmargin=5pt,
    framerule=0.5pt,
    linenos
}

\newminted[JavaSourceCode]{java}{
    labelposition=topline,
    frame=single,
    numbersep=2pt,
    fontsize=\tiny,
    xleftmargin=5pt,
    framerule=0.5pt,
    linenos
}

\definecolor{magnolia}{rgb}{0.97, 0.96, 1.0}
\definecolor{shadecolor}{rgb}{0.97, 0.96, 1.0}

\newmintinline[codePython]{python}{fontsize=\small}
\newmintinline[codeJava]{java}{fontsize=\small}
\newmintinline[snippetJava]{java}{fontsize=\small,bgcolor=magnolia}
\newmintinline[snippetPython]{python}{fontsize=\small,bgcolor=magnolia}
\newmintinline[snippetPerl]{perl}{fontsize=\small,bgcolor=magnolia}

\setlist{nosep, topsep=0pt, partopsep=0pt, parsep=0pt, itemsep=0pt}

\setlist[itemize]{noitemsep, topsep=0pt, leftmargin=*}

\journal{Information and Software Technology}

\begin{document}

\begin{frontmatter}

\title{Security in the Age of AI Teammates: An Empirical Study of Agentic Pull Requests on GitHub}

\author{Mohammed Latif Siddiq, Xinye Zhao, Vinicius Carvalho Lopes, Beatrice Casey, Joanna C. S. Santos} \affiliation{organization={Department of Computer Science and Engineering},addressline={University of Notre Dame}, 
            city={Notre Dame},
            postcode={46556}, 
            state={IN},
            country={USA}}

\begin{abstract}
\textbf{Context.}
Autonomous coding agents are increasingly deployed as AI teammates in modern software engineering, independently authoring pull requests (PRs) that modify production code at scale. Prior empirical studies have primarily examined their productivity and acceptance rates, leaving the role of autonomous agents in software security and the dynamics of human review largely unexplored.

\textbf{Objectives.}
This study aims to systematically characterize how autonomous coding agents contribute to software security in practice, how these security-related contributions are reviewed and accepted, and which observable signals are associated with PR rejection.

\textbf{Methods.}
We conduct a large-scale empirical analysis of agent-authored PRs using the AIDev dataset, comprising over 33,000 curated PRs from popular GitHub repositories. Security-relevant PRs are identified using a keyword filtering strategy, followed by manual validation, resulting in 1,293 confirmed security-related agentic PRs. We then analyze prevalence, acceptance outcomes, and review latency across autonomous agents, programming ecosystems, and types of code changes. Moreover, we apply qualitative open coding to identify recurring security-related actions and underlying intents, and examine review metadata to identify early signals associated with PR rejection.

\textbf{Results.}
Security-related Agentic-PRs constitute a meaningful share of agent activity (approximately 4\%). Rather than focusing solely on vulnerability fixes, agents most frequently perform supportive security hardening activities, including testing, documentation, configuration, and improved error handling. Compared to non-security PRs, security-related Agentic-PRs exhibit lower merge rates and longer review latency, reflecting heightened human scrutiny, with variation across agents and programming ecosystems. PR rejection is more strongly associated with complexity and verbosity than with explicit security topics, and trained text-based models outperform both structured-feature and prompting-based LLM baselines.

\textbf{Conclusion.}
Autonomous coding agents already perform a non-trivial amount of security-relevant work in real-world repositories, but these contributions are reviewed more cautiously by human maintainers. Security review of agent-authored code extends beyond vulnerability content alone and is shaped by contextual and
cognitive factors.

\end{abstract}


\end{frontmatter}
\section{Introduction}

\subsection{AI Assistants in Modern Software Engineering}
Large Language Models (LLMs) are increasingly prevalent in everyday applications and assistive tasks, driven by advances in transformer-based architectures and large-scale training~\cite{brown2020language,ouyang2022training,openai2023gpt4,attention2017}. These advances have enabled models with unprecedented scale (often comprising billions or even trillions of parameters) and strong capabilities across a wide range of natural language and code-related tasks~\cite{openai2023gpt4}. As a result, software engineering (SE) is entering a new phase increasingly shaped by \emph{AI teammates}: autonomous, task-driven agents capable of performing complex software development activities such as feature implementation, debugging, testing, and code review with limited human oversight~\cite{li2025se30,hassan2025agenticse}.

Autonomous coding agents are now being deployed in modern software engineering workflows, where they independently author pull requests (PRs) that modify production code at scale. Recent progress in LLM reasoning, tool use, and autonomous planning has enabled such agents to perform multi-step development tasks, interact with version control systems, and submit PRs with minimal human intervention~\cite{yao2023reactsynergizingreasoningacting,zhou2024webarenarealisticwebenvironment}. Consequently, agent-authored PRs are no longer confined to experimental demonstrations but are increasingly observed in real-world GitHub repositories and open-source projects~\cite{watanabe2025agenticpr,jie2025domain}.

Prior empirical research on LLMs and autonomous agents in software engineering has largely focused on productivity-oriented outcomes, including code generation quality, task completion rates, acceptance ratios, and developer efficiency~\cite{watanabe2025agenticpr,chen2021evaluatinglargelanguagemodels,zhang2024surveylargelanguagemodels}. These studies demonstrate that autonomous agents can accelerate development workflows and contribute meaningfully to large codebases. However, productivity alone is an incomplete proxy for trustworthiness, particularly in security-critical contexts, where subtle defects may introduce latent vulnerabilities rather than immediate functional failures~\cite{pearce2022asleep,siddiq2022empirical,allegrini2025formalizing,moshkovich2025uncertainty}.

\subsection{Challenges of Agent-Authored Pull Requests}
Software security poses distinct challenges for autonomous coding agents. Security-relevant code changes often require nuanced reasoning about threat models, privilege boundaries, cryptographic usage, configuration semantics, and backward compatibility, areas in which prior work has shown that LLMs may exhibit inconsistent understanding, brittle reasoning, or overconfidence~\cite{pearce2022asleep,li2024knowledge}. Errors in such changes may evade test suites and static analysis tools, weakening a system’s security posture without immediately observable failures. Despite these risks, the role of autonomous coding agents in security-relevant software development remains poorly understood.

\subsection{Limitations of Prior Studies}
Existing empirical studies provide limited insight into this problem space. Most analyses treat agent-authored pull requests as a largely homogeneous category, without distinguishing security-relevant changes from routine maintenance tasks such as refactoring, dependency updates, or formatting fixes~\cite{zhang2024surveylargelanguagemodels,watanabe2025agenticpr,wang2025automated}. At the same time, research on LLM-based security evaluation has primarily focused on isolated tasks, such as vulnerability detection, secure code generation, formal verification, or question answering, rather than on how security work unfolds within realistic development workflows and human review processes~\cite{pearce2022asleep,siddiq2024sallm,chatterjee2025proofwright}. While these efforts help understand models' capabilities, they offer limited visibility into how autonomous agents contribute to security in practice and how such contributions are scrutinized by human developers.

Consequently, we lack a systematic understanding of (i) how frequently autonomous coding agents engage in security-relevant development, (ii) what types of security work they perform in practice, (iii) the actions and intents reflected in agent-authored security changes, and (iv) how human developers evaluate and scrutinize such work during code review. Addressing these gaps is essential for assessing the trustworthiness and governance of GenAI-enabled software systems as autonomous coding agents become increasingly embedded in real-world development workflows.

In this work, we present the first \textbf{large-scale empirical study of \emph{security-related Agentic pull requests} (Agentic-PRs) in real-world GenAI-enabled GitHub repositories}. To conduct this study, we used the AIDev dataset~\cite{li2025aidev}, which contains a curated dataset of 33,596 PRs authored by five widely deployed autonomous coding agents across diverse repositories, programming languages, and software ecosystems. From this corpus, we systematically identified \emph{security-related Agentic-PRs} using a rule-based filtering strategy applied to PR titles and descriptions, and then manually vetted them. 
We analyze each security-related Agentic-PR along multiple complementary dimensions. First, we measure \emph{prevalence}, quantifying how frequently autonomous agents engage in security-relevant development relative to their overall activity. Second, we examine \emph{security outcomes} by analyzing merge and rejection patterns across agents, ecosystems, and types of security-related changes. Third, we analyze the security \emph{actions} taken by the agents and their \emph{intents} behind them. Fourth, we study \emph{human review behavior} by analyzing review latency and merge decisions as conservative, outcome-based proxies for reviewer scrutiny in security-related Agentic-PRs, and comparing them with those in human-authored PRs. Finally, we investigate \emph{early indicators of perceived risk} by analyzing lightweight, early-available signals, such as PR size, title, and description length, and security-sensitive terminology, to assess which characteristics are associated with PR rejection.

Throughout the study, we employ a rigorous analysis process that combines quantitative measurement with qualitative interpretation. In particular, we complement large-scale statistical analyses with open coding of a manually validated subset of security-related Agentic-PRs to characterize recurring security actions and underlying intents. Disagreements in qualitative coding are resolved through discussion to ensure interpretive consistency. Together, this methodology enables us to systematically characterize how autonomous coding agents engage with security-critical development tasks in practice and how human developers evaluate and scrutinize such contributions during real-world code review.

\subsection{Manuscript Organization}

The remainder of this manuscript is organized as follows. Section~\ref{sec:background} introduces the necessary background and situates our work within the existing literature on autonomous coding agents, pull requests, and software security. Section~\ref{sec:methodology} describes our empirical study design, including the dataset, security relevance identification, and analysis procedures. Section~\ref{sec:results} presents the results addressing each research question. Section~\ref{sec:discussion} discusses the implications of our findings, limitations, and directions for future work. Finally, Section~\ref{sec:conclusion} concludes the paper.

 \section{Background \& Related Work}\label{sec:background}

This section introduces the background necessary to contextualize our study and positions it with respect to prior work on LLMs in software engineering, software security evaluation, and empirical studies of pull requests and code review.
\subsection{LLMs and Autonomous Coding Agents in Software Engineering}

Large Language Models (LLMs) have rapidly become integral to modern software engineering workflows, supporting tasks such as code completion, code generation, debugging, testing, and documentation~\cite{chen2021evaluatinglargelanguagemodels,zhang2024surveylargelanguagemodels}. Recent advances in model scale, reasoning, and tool use have enabled the emergence of \emph{\textbf{autonomous coding agents}}: systems that can independently perform multi-step development tasks, interact with development tools and version control systems, and author PRs with limited or no human intervention~\cite{yao2023reactsynergizingreasoningacting,zhou2024webarenarealisticwebenvironment}.

Early empirical studies of LLM-authored artifacts have primarily emphasized productivity-oriented outcomes, such as task completion rates, merge or acceptance ratios, and developer efficiency~\cite{chen2021evaluatinglargelanguagemodels,zhang2024surveylargelanguagemodels}. While these results demonstrate that autonomous agents can meaningfully contribute to real-world repositories, they provide limited insight into how such contributions affect \emph{non-functional properties}, particularly software security. Our work builds on this line of research by explicitly examining agent-authored PRs through a security-specific and workflow-centric lens.

Several works have established conceptual foundations for agentic software engineering. Hassan \etal~\cite{hassan2025agenticse} articulate the notion of \emph{Software Engineering~3.0}, positioning autonomous AI agents as collaborative teammates and outlining foundational pillars and open research challenges. Complementing this perspective, Hoda~\cite{hoda2025beyondcode} argues that agentic software engineering must extend beyond code to encompass shared vision, values, and vocabulary, while Sapkota and Roumeliotis~\cite{sapkota2025vibecoding} contrast agentic coding with more ad-hoc ``vibe coding'' practices. Earlier work in agent-oriented software engineering~\cite{zambonelli2004ao} provides historical context for many of the coordination, autonomy, and governance challenges that now re-emerge in LLM-based agentic systems.

Beyond conceptual framing, multiple studies propose methodologies and platforms for agent-based software development. Bandara \etal~\cite{bandara2025agentsway} introduce \emph{Agentsway}, a development methodology for AI-agent-based software teams, while Rasheed \etal~\cite{rasheed2024codepori} and Sami \etal~\cite{sami2024unified} present large-scale multi-agent platforms that support autonomous software development. Surveys by Wang \etal~\cite{wang2025agenticprogramming} and Li \etal~\cite{li2025se30} further synthesize techniques, challenges, and opportunities in agentic programming and AI teammates, highlighting issues of coordination, evaluation, and reliability.

Understanding how humans collaborate with autonomous agents is also critical for their adoption in practice. Klieger \etal~\cite{klieger2024chatcollab} empirically study collaboration patterns between humans and AI agents in software teams, while Ronanki~\cite{ronanki2025trustworthy} focuses on mechanisms for fostering trustworthy human--agent collaboration in LLM-based multi-agent systems. Practitioner-oriented studies reinforce these socio-technical concerns, emphasizing trust calibration, accountability, and workflow disruption across the software development life cycle~\cite{akbar2025practitioner,sarkar2025productivity}.

Empirical evidence on the behavior of agentic systems in real-world software artifacts is rapidly accumulating. Watanabe \etal~\cite{watanabe2025agenticpr} analyze Claude Code PRs across 157 open-source projects and compare them with human-authored PRs. They report that agent-authored PRs are accepted less often than human PRs (83.8\% vs.\ 91.0\%), although the share merged without revision is similar across the two groups (54.9\% vs.\ 58.5\%). Complementing this smaller-scale comparison, recent large-scale studies further show that agent-authored PR outcomes vary substantially across agents and software domains~\cite{jie2025domain,watanabe2025agenticpr}. Related work also suggests that integration outcomes depend not only on code correctness, but also on review and quality-assurance practices in agentic software ecosystems~\cite{hasan2025testing}.

In contrast to these studies, our work does not focus on general productivity or domain-level acceptance patterns of agent-authored PRs; instead, we explicitly center on security-relevant Agentic-PRs and analyze the security intents, actions, and review scrutiny that emerge when autonomous agents engage in security-sensitive software changes, dimensions not examined in prior work.

Beyond productivity and collaboration, several studies examine correctness, safety, and reliability properties of agentic systems. Allegrini \etal~\cite{allegrini2025formalizing} formalize safety, security, and functional properties of agentic AI systems, while Chatterjee \etal~\cite{chatterjee2025proofwright} introduce agentic approaches to formal verification for CUDA programs. Moshkovich and Zeltyn~\cite{moshkovich2025uncertainty} further explore techniques for observing, analyzing, and optimizing agentic systems under uncertainty, underscoring the need for principled governance and monitoring mechanisms.

While prior research has advanced conceptual frameworks, development methodologies, and empirical analyses of agentic software artifacts, most studies either focus on productivity and collaboration or analyze agentic outputs at a coarse granularity. In contrast, our work specifically examines \emph{security-relevant behavior} of autonomous coding agents within real-world pull-request workflows, integrating large-scale quantitative analysis with qualitative interpretation of agent actions, intents, and human review dynamics.

\subsection{LLMs and Software Security}

A growing body of work has examined the security implications of LLM-generated code. Early studies found that a substantial fraction of LLM-generated programs contain vulnerabilities, even when models produce syntactically correct and functional code \cite{pearce2022asleep}. Subsequent empirical analyses further document security weaknesses across languages and tasks \cite{siddiq2022empirical,sandoval2022security,hajipour2023systematically}, and propose mitigation strategies such as static-analysis–guided ranking or filtering of model outputs \cite{siddiq2023franc}. 

To support systematic security evaluation of LLM-generated code, several benchmarks and frameworks have been proposed. These include scenario-based vulnerability assessments \cite{pearce2022asleep}, CWE-driven prompt suites such as SecurityEval \cite{siddiq2022seceval}, and static-analysis–based testing pipelines such as \texttt{CyberSecEval} \cite{bhatt2023purple}. While these efforts advance the evaluation of LLMs for secure code generation and vulnerability detection, they largely focus on isolated code-level tasks. In contrast, our study investigates security behavior in a \emph{workflow-centric setting}, analyzing how autonomous agents perform security-related work within real PR-based development processes.

\subsection{PRs and Security}

Pull requests (PRs) are a central coordination mechanism in modern software development, serving as a primary vehicle for code contribution as well as a focal point for review, discussion, and quality assurance. Several empirical studies have examined PR processes, identifying factors that influence acceptance and rejection decisions, review latency, and reviewer behavior. Prior studies show that PR outcomes are shaped by a combination of technical characteristics, such as code complexity, change size, and test coverage, and social or contextual factors, including contributor experience and reputation, reviewer workload, and social signals embedded in discussion threads~\cite{mcintosh2014impact,rigby2014peer,baysal2016investigating}. Subsequent work further demonstrates that review difficulty and latency are influenced by how changes are scoped and presented, with large or complex PRs being harder to review and more likely to experience delays or rejection~\cite{ram2018easierreview,zhang2022prlatency}. 

\emph{Security-related pull requests} introduce additional challenges beyond those observed for general code changes, as they often involve subtle design decisions, domain-specific expertise, and risk-sensitive trade-offs. In this work, we define a \textbf{security-related PR} as a pull request whose stated intent or code changes explicitly involve security-relevant concerns, such as vulnerability mitigation, authentication or authorization logic, cryptographic usage, secure configuration, dependency security updates, or compliance with security standards and regulations. This definition is consistent with prior empirical security studies that operationalize security relevance using a combination of developer intent, textual artifacts (\eg~PR titles and descriptions), and the nature of the modified code~\cite{rigby2014peer,wang2025automated}.

Empirical studies suggest that security-related PRs are treated differently from general PRs during review. Lenarduzzi \etal~\cite{lenarduzzi2021codequalitypr} show that PRs containing quality issues, including security-related problems, are less likely to be accepted, while Zhang \etal~\cite{zhang2021prdecision} highlight that decision-making criteria vary substantially depending on the perceived risk and impact of the change. More recent work has examined security PRs in the context of automated dependency management tools. Rebatchi \etal~\cite{rebatchi2024dependabotsecurityprs} conduct a large-scale empirical study of Dependabot security PRs and show that, despite being automated and security-focused, such PRs still face non-trivial review delays and rejection rates, reflecting developer caution toward security updates. These findings reinforce the notion that security-related PRs often receive heightened scrutiny due to their potential to introduce subtle but severe vulnerabilities. Complementary work \cite{wang2025automated} compares AI-generated and human security patches using the AIDev dataset and examines domain-level variation in agentic pull-request acceptance; in contrast, our study centers security as a first-class dimension by examining the specific security actions agents perform, the intents underlying those actions, how human reviewers scrutinize security-relevant Agentic-PRs, and what early risk signals emerge during review.

Despite this growing body of work, existing PR research has largely focused on human-authored contributions or on automated tools operating in narrowly scoped settings, such as dependency updates. At the same time, recent studies of agent-authored PRs tend to treat security as a secondary concern or subsume it under general productivity metrics. As autonomous coding agents increasingly submit PRs that touch security-sensitive code paths, understanding how security-related agent-authored PRs are evaluated, scrutinized, and governed within real-world review workflows remains an open and underexplored problem which is addressed in this current work.
 \section{Methodology}\label{sec:methodology}
\begin{figure}[!htbp]
    \centering
    \includegraphics[width=\linewidth]{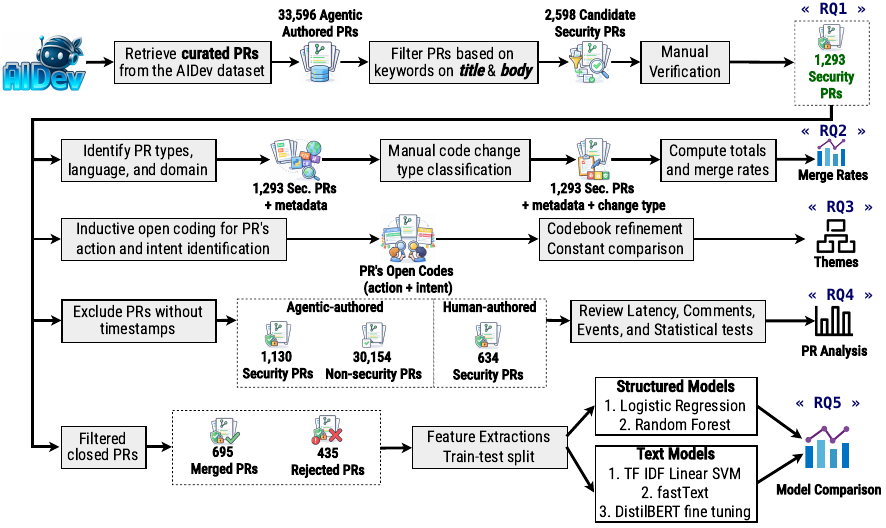}
    \caption{Overview of our Study's Methodology.}
    \label{fig:overview}
\end{figure}

This section describes the methodology followed to answer our RQs. As shown in Figure~\ref{fig:overview}, our work follows a structured, multi-stage empirical process. First, we identify security-relevant Agentic-PRs through keyword-based filtering followed by manual validation. Second, we conduct large-scale quantitative analyses to examine the prevalence of security-related contributions, their outcomes, and patterns of human review and scrutiny. Finally, we perform qualitative open coding and predictive modeling to analyze agent security actions, design intents, and early indicators of risk or rejection.

\subsection{Research Questions (RQs)}

Autonomous coding agents can author pull requests that modify production code, including changes with potential security implications. In the following research questions (RQs), we investigate the prevalence, outcomes, and oversight of security-related agent activity in practice.

\rques{\textbf{RQ1}: How frequently do autonomous coding agents contribute security-relevant software changes to GitHub repositories?}

This RQ characterizes how often autonomous coding agents engage in security-critical development activities. By analyzing the frequency and distribution of security-relevant PRs across agents, we establish a system-level baseline of agent participation in security-sensitive software engineering tasks.

\rques{\textbf{RQ2}: How do the outcomes of agent-authored security-related PRs vary across agents, programming ecosystems, and code change types?}

This research question investigates how security-related PR outcomes differ across autonomous coding agents, programming ecosystems (\eg languages and domains), and categories of code changes. By comparing merge rates and review outcomes across these dimensions, we assess how reliably agents contribute security-relevant changes in different technical contexts.

\rques{\textbf{RQ3}: What security-related actions and design intents are reflected in agent-authored security PRs?}

This RQ examines how autonomous coding agents engage with security in practice. Using open coding, we analyze agent-authored security-relevant pull requests, including titles, descriptions, and code changes, to identify recurring security actions and the underlying intents that motivate them. This analysis provides insight into the patterns through which agents attempt to improve or modify system security.

\rques{\textbf{RQ4}: How does human review shape the outcomes of agent-authored security PRs?}

This research question examines how human reviewers respond to security-relevant contributions produced by autonomous coding agents. By analyzing merge outcomes and review latency as conservative proxies for reviewer scrutiny, we assess whether security-related agent-authored PRs receive heightened, differentiated scrutiny and how this scrutiny varies across agents and repositories. In addition, we compare them with human-authored security PRs from the same repositories.

\rques{\textbf{RQ5}: Which early, observable signals help identify potentially risky or rejected security-related agent-authored pull requests?}

This research question investigates which early signals available at PR creation time are associated with rejection or perceived risk. By analyzing metadata and textual characteristics, we identify factors that may serve as lightweight indicators for risk assessment and governance of security-relevant contributions produced by autonomous coding agents.

\subsection{Dataset of Pull Requests Authored by Autonomous Coding Agents}
\label{subsec:dataset}
To answer our RQs, we need a dataset of AI-authored PRs that supports fine-grained analysis of security-related development and human oversight. To this end, we use the \textbf{AIDev} dataset~\cite{li2025aidev},  a large-scale corpus containing  {932,791 PRs} authored by autonomous coding agents in real-world GitHub repositories.  AIDev identifies agent-authored PRs through agent-specific bot accounts (\eg~\texttt{copilot-swe-agent[bot]}), PR labels injected by agent platforms, and commit metadata indicating non-human authorship.
The AIDev dataset also provides a \textit{curated subset} of {33,596 agentic-authored PRs} drawn from {2,807 GitHub repositories} with at least {100} stars, covering five widely deployed agents: OpenAI Codex (21,799 PRs),
Copilot (4,970), Devin (4,827), Cursor (1,541), and Claude Code (459). This curated subset of PRs includes enriched artifacts such as PR \textit{titles}, \textit{descriptions}, \textit{timestamps}, \textit{merge outcomes}, \textit{review comments}, \textit{review decisions}, \textit{commit metadata}, and \textit{file-level diffs}. The five agents differ in how they are integrated into development workflows, which shapes the types of tasks they address. Copilot operates as an issue-to-PR pipeline: maintainers assign a GitHub issue to the agent, which autonomously implements and opens the PR. Devin can independently explore a codebase and select tasks via a Slack or web interface. OpenAI Codex, Cursor, and Claude Code are invoked directly by developers via APIs, IDEs, or terminals.

While the full dataset enables broad coverage, it does not uniformly provide the artifacts required to study security outcomes and human review behavior. Therefore, we use AIDev's \textbf{curated subset of 33,596 PRs}.

\subsection{Answering RQ1: Prevalence of Security-Relevant Agentic-PRs}
\label{subsec:rq1_method}

To answer \textbf{RQ1}, we quantify the prevalence of security-relevant pull requests authored by autonomous coding agents. This analysis establishes a system-level baseline of how frequently GenAI-enabled software systems engage in security-critical development activities.
To do so, we identify security-related Agentic-PRs through a two-stage process that combines automated filtering with manual validation.

\subsubsection{Stage 1: Keyword-based PR filtering}

In the first stage, we used a conservative, keyword-based filter to identify \emph{candidate} security-related PRs based on their titles and descriptions. This step is intentionally inclusive and designed to capture PRs that \emph{may} involve security concerns. Specifically, we searched for a curated list of security-related terms commonly used in software development practice, spanning multiple security dimensions. These terms span six categories (Table~\ref{tab:keywords}):
(i) general security concepts (\ie~\textit{security}, \textit{vulnerability}, \textit{threat}, \textit{attack});
(ii) identity and access management (\ie~\textit{auth}, \textit{authentication}, \textit{authorization}, \textit{password}, \textit{credential}, \textit{secret}, \textit{token});
(iii) cryptography (\ie~\textit{encryption});
(iv) attack vectors (\ie~\textit{xss}, \textit{csrf}, \textit{injection}, \textit{exploit}, \textit{cve}, \textit{buffer overflow}, \textit{rce}, \textit{remote code execution}, \textit{privilege escalation}, \textit{malicious});
(v) privacy and compliance (\ie~\textit{pii}, \textit{leaks}, \textit{gdpr}, \textit{hipaa}, \textit{audit}, \textit{compliance}); and
(vi) defensive operations (\ie~\textit{patch}, \textit{sanitize}, \textit{denial of service}, \textit{dos}, \textit{ddos}).

\begin{table}[t]
\centering
\caption{Security keyword taxonomy used for PR candidate selection.}
\label{tab:keywords}
\begin{tabular}{ll}
\toprule
\textbf{Category} & \textbf{Keywords} \\
\midrule
General Security
  & security, vulnerability, threat, attack \\
\midrule
Identity \& Access Management
  & auth, authentication, authorization, \\
  & password, credential, secret, token \\
\midrule
Cryptography
  & encryption \\
\midrule
Attack Vectors
  & xss, csrf, injection, exploit, cve, \\
  & buffer overflow, rce, remote code execution, \\
  & privilege escalation, malicious \\
\midrule
Privacy \& Compliance
  & pii, leaks, gdpr, hipaa, audit, compliance \\
\midrule
Defensive Operations
  & patch, sanitize, denial of service, dos, ddos \\
\bottomrule
\end{tabular}
\end{table}
 
These keywords are matched using case-insensitive regular expressions with word-boundary constraints to reduce false positives. A PR is then flagged as a \emph{candidate security PR} if at least one keyword appears in its title or body. After this filtering step, we initially identify \textbf{2,598 candidate security-related agentic-authored PRs} in the curated AIDev subset.

\subsubsection{Stage 2: Manual Vetting}

In the second stage, we manually inspect the 2,598 candidate PRs to determine whether they are \emph{genuinely security-relevant}. Two authors with 4–6 years of experience in software development and software security research independently performed this manual analysis. We deliberately rely on human expert validation rather than LLM-based labeling in this stage. While LLMs could have provided an additional source of analysis during candidate selection or annotation, we use human validation to establish the final security-relevance labels and avoid introducing model-dependent variability into the ground truth \cite{schroeder2025just}.

To assess annotation reliability, we first sampled 335 PRs from the \textbf{2,598 candidate security-related agentic-authored PRs}, with a 95\% confidence level and a 5\% margin of error. Both annotators independently labeled these sampled PRs, and we measured inter-rater agreement using Cohen’s $\kappa$ coefficient~\cite{cohen1960coefficient}. The resulting $\kappa$ score of \textbf{0.91} indicates \emph{almost perfect agreement}, demonstrating a high level of annotation consistency.

A \textbf{\textit{security PR}} is defined as a PR whose primary intent is to \emph{prevent, mitigate, detect, or remediate a security vulnerability or security risk}. This includes fixes for known vulnerabilities, hardening of security mechanisms, corrections to authentication or authorization logic, cryptographic updates, and changes explicitly motivated by security or compliance concerns. PRs that merely mention security-related terms incidentally (\eg~documentation updates, configuration defaults, or unrelated refactoring) are treated as false positives. Based on this manual validation process, we identified \textbf{1,293 security-relevant Agentic-PRs}, giving an overall keyword-filter precision of \textbf{49.8\%}. Precision varies substantially by keyword: terms such as \texttt{CVE}, \texttt{vulnerability}, \texttt{malicious}, and \texttt{encryption} yield 100\% precision (every matched PR was confirmed), while broader terms such as \texttt{patch}, \texttt{compliance}, and \texttt{token} yield lower precision (13--37\%), reflecting their frequent incidental use in non-security contexts. This variation motivates the manual validation stage, meaning the keyword filter is treated as a high-recall first pass, with precision enforced
through human review.

\subsection{Answering RQ2: Differences Across Agents, Ecosystems, and Code Change Types}
\label{subsec:rq2_method}

To answer \textbf{RQ2}, we analyze how the outcomes of Agentic-PRs vary across autonomous coding agents, project domains, programming ecosystems, and types of code changes. We group security-relevant Agentic-PRs by coding agent and compute, for each agent, the total number of security PRs, the number of closed PRs, and the corresponding merge rate. To provide additional context, we also identify the dominant programming language associated with each agent’s security PRs. We likewise avoid LLM-assisted categorization in this stage, so that the resulting security change labels remain grounded in human expert judgment rather than in additional model-mediated interpretation.

Project domain is derived automatically from the \texttt{language} field of the AIDev dataset, which records the primary programming language of each repository as reported by GitHub. Languages are mapped to five coarse domains using a deterministic lookup: \emph{Web} (TypeScript, JavaScript, HTML, CSS, PHP, Ruby), \emph{Systems} (C, C\texttt{++}, Rust, Go), \emph{Data/ML} (Python, Jupyter Notebook), \emph{Enterprise} (Java, C\#), and \emph{Ops} (Shell, Dockerfile, Makefile). Repositories whose primary language does not fall into any of these groups are labeled \emph{Other}. We then manually checked the resulting language and domain assignments for consistency.

We further examine how security outcomes vary across different types of code changes.
As a starting point, we leverage the coarse-grained PR-type tags provided by the \textbf{AIDev dataset}~\cite{li2025aidev}, which are automatically inferred using lightweight, rule-based heuristics applied to PR titles (\eg~\texttt{fix}, \texttt{feat}, \texttt{add}, \texttt{config}, \etc).
As noted by Li et al.~\cite{li2025aidev}, these tags are intended to provide an approximate characterization of PR intent rather than definitive ground truth and may exhibit overlap or ambiguity, particularly for security-related changes.

To address these limitations, we manually inspected all \textbf{1,293 security-relevant Agentic-PRs} identified in RQ1 (Section~\ref{subsec:rq1_method}). During this process, we considered both the PR content (code diffs, descriptions, and discussion) and the AIDev-provided PR-type tags as contextual cues. Rather than relying on individual fine-grained labels (\eg~\texttt{fix} vs.\ \texttt{patch} vs.\ \texttt{resolve}), we assigned each PR to a single \emph{coarse-grained security change category} that best reflects the primary nature of the security-related modification. This manual categorization follows the same validation protocol used for security relevance labeling in RQ1. We assign each PR to the security change category that best reflects the \emph{primary} nature of the modification. In the vast majority of cases, this assignment is unambiguous: \textbf{98.9\%} (1,279 of 1,293) of confirmed security PRs were assigned a single category by all coders without disagreement. The remaining \textbf{1.1\%} (14 PRs) address concerns spanning two categories simultaneously (\eg~a PR that upgrades a vulnerable dependency \emph{and} patches application logic, or one that refactors authentication code \emph{and} updates a compliance configuration). For these borderline cases, coders assigned both applicable categories rather than imposing an artificial single label, and the PR is counted in both categories for frequency analysis but excluded from merge-rate comparisons that require a single-group assignment. The four categories are:

\begin{itemize}
  \item \textbf{\textit{Dependency Update}}: PRs that modify third-party dependencies or library versions without directly changing application logic. These changes are typically preventive or maintenance-oriented and may incorporate upstream security fixes. These PRs are frequently flagged with \texttt{bump}, \texttt{upgrade}, \texttt{dependency}, and \texttt{chore(deps)} tags.

  \item \textbf{\textit{Vulnerability Fix}}: PRs that directly remediate identified security flaws or weaknesses in the codebase, including fixes associated with known vulnerabilities, CVEs, or exploitable behaviors. Importantly, CVE assignment is \emph{not} a requirement for this category: a PR is classified here whenever its primary purpose is to remediate a concrete security flaw, regardless of whether a CVE identifier is referenced, since not all vulnerabilities in open-source repositories receive a CVE assignment~\cite{simion2026beyond,ayala2025mixed,mu2018crowd}. These PRs are commonly flagged with \texttt{fix}, \texttt{patch}, \texttt{resolve}, \texttt{vuln}, \texttt{cve}, and \texttt{exploit} tags.

  \item \textbf{\textit{Security Feature}}: PRs that introduce new security-related functionality or extend existing security mechanisms, such as authentication, authorization, access control, encryption, or auditing logic. 
These PRs are frequently flagged with  \texttt{feat}, \texttt{add}, \texttt{implement}, \texttt{support}, and \texttt{new} tags.

  \item \textbf{\textit{Config / Compliance}}: PRs that adjust security-relevant configuration files, policy definitions, or compliance-related artifacts (\eg access rules, security settings, or regulatory constraints) without introducing new application logic. 
These PRs are frequently flagged with \texttt{config}, \texttt{setting}, \texttt{policy}, \texttt{audit}, and \texttt{compliance} tags.
\end{itemize}

Using these manually validated coarse-grained categories, we compute the volume and merge rate of security-related Agentic-PRs across agents, ecosystems, and change types. This approach enables systematic comparison of security outcomes while avoiding reliance on noisy or overlapping fine-grained PR labels.

\subsection{Answering RQ3: Security Actions and Intents in Agentic-PRs}
\label{subsec:rq3_method}

To answer \textbf{RQ3}, we conduct a qualitative analysis of security-relevant pull requests authored by autonomous coding agents using \emph{open coding} \cite{strauss1998basics}. Our goal is to understand \emph{what security-relevant Agentic-PRs are doing in practice}, rather than to evaluate their correctness or downstream impact.

For each  of the 1,293 agentic security PRs we collected in RQ1, we analyze their title and description, which often articulate the agent’s stated intent. We apply open coding following established qualitative analysis practices. Two authors independently examine PR artifacts line by line and assign two complementary types of codes: 
(1) \emph{security-related actions}, which capture \textbf{what concrete change the agent performs in the code} (\eg~adding input validation, modifying authentication logic, introducing access controls, or updating dependencies); and 
(2) \emph{security intents}, which capture \textbf{why the change is being made}, reflecting the agent’s stated or implied motivation (\eg~mitigating a vulnerability, hardening security posture, preventing misuse, or improving compliance).

Actions and intents are coded separately to avoid conflating implementation mechanisms with their underlying security rationale. Codes are derived inductively from the data without relying on predefined vulnerability taxonomies such as Common Weakness Enumeration (CWE) or the Open Worldwide Application Security Project (OWASP).

\begin{figure}
    \centering
    \includegraphics[width=\linewidth]{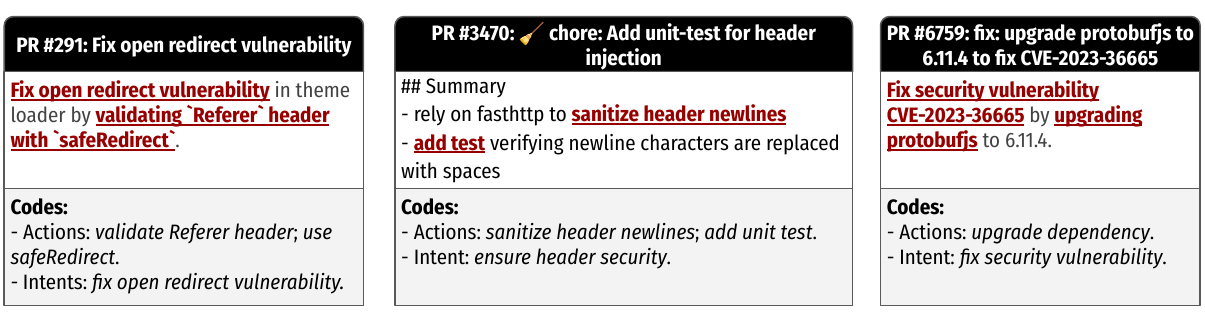}
    \caption{
    Examples of open coding applied to security-related Agentic pull requests. 
    }
    \label{fig:AgenticOpenCoding}
\end{figure}

To better illustrate our qualitative analysis process, Figure~\ref{fig:AgenticOpenCoding} presents three representative security-related Agentic-PRs annotated using open coding.
In each example, salient security-relevant elements in the PR title and description are highlighted (shown in red) and mapped to corresponding \emph{action} and \emph{intent} codes.

For instance, in PR~\#291, which addresses an open redirect vulnerability, phrases such as “validating the \texttt{Referer} header” and “using \texttt{safeRedirect}” are highlighted and coded as \emph{Actions}, capturing the concrete remediation steps taken by the agent.
The stated rationale, “fix open redirect vulnerability”, is separately coded as the \emph{Intent}, reflecting the underlying security motivation rather than the implementation mechanism. Similarly, PR~\#3470 focuses on preventing header injection by sanitizing newline characters and adding unit tests. In this example, the introduction of sanitization logic and test cases is coded as \emph{Actions}, while the broader goal of ensuring header security is captured as the \emph{Intent}. In contrast, PR~\#6759 addresses CVE-2023-36665 by upgrading the \texttt{protobufjs} dependency to a secure version. The dependency upgrade is coded as the primary \emph{Action}, whereas mitigating a known vulnerability constitutes the associated \emph{Intent}.

Throughout the process, we employ constant comparison, iteratively refining codes as new PRs are analyzed. Conceptually overlapping codes are merged, while distinct patterns are preserved to maintain analytic granularity. Analytic memos are maintained to document coding decisions and emerging themes. Disagreements between coders are resolved through discussion and consensus. The final codebook is then applied consistently across the dataset. 

While our goal is not statistical generalization, this process ensures interpretive rigor and transparency. Using the finalized codes, we report the frequency and distribution of recurring security actions and intents across agents and programming ecosystems.

\subsection{Answering RQ4: Reviewer Behavior and Scrutiny of Security Agentic-PRs}
\label{subsec:rq4_method}

To address \textbf{RQ4}, we examine whether human reviewers treat security-relevant Agentic-PRs differently from non-security Agentic-PRs. Since reviewer intent and cognitive effort are not directly observable, we operationalize \emph{reviewer scrutiny} using a set of conservative, outcome- and interaction-based proxies that have been widely adopted in empirical software engineering research~\cite{li2025aidev}.

\paragraph{Review Scrutiny Proxies}

We characterize reviewer scrutiny using multiple complementary signals that capture both the \emph{duration} and \emph{intensity} of the review process:

\begin{itemize}[ leftmargin=*]
    \item \textbf{Review Latency}. We measure review latency as the elapsed time (in hours) between PR creation and PR closure (merged or rejected). Longer latencies may indicate more careful inspection, extended discussion, or reviewer hesitation, whereas shorter latencies may reflect expedited handling. PRs with missing timestamps or non-positive durations are excluded.
    
    \item \textbf{Merge Rate}. We compute the merge rate as the proportion of closed PRs that were merged. Lower merge rates may indicate stricter evaluation criteria or higher reviewer skepticism toward the proposed changes.
    
    \item \textbf{Review Comments}. We measure the total number of comments posted on a PR, capturing both reviewer questions and discussion depth.
    
    \item \textbf{Review Events}. We count the total number of formal review actions (\eg~\emph{Approved}, \emph{Changes Requested}, \emph{Commented}) associated with a PR. We additionally analyze the distribution of event types to distinguish between passive acknowledgment and explicit corrective feedback.
\end{itemize}

\paragraph{Control Variables: PR Complexity and Effort}

To disentangle reviewer scrutiny from potential confounding effects caused by PR size or complexity, we additionally measure several control variables:

\begin{itemize}[ leftmargin=*]
    \item \textbf{Code Churn}, defined as the total number of lines added and deleted in the PR.
    \item \textbf{Files Touched}, measured as the number of distinct files modified.
    \item \textbf{Description Length}, computed as the word count of the PR body.
\end{itemize}
PR size is operationalized as the word count of the PR description body, the only size-related signal available uniformly across all PRs in the dataset. Control is achieved through \emph{tertile stratification}, meaning all closed PRs are partitioned into three equal-frequency groups (small: $<$36 words; medium: 36--69 words; large: $>$69 words), and merge-rate comparisons are repeated independently within each stratum.

For all closed PRs in the curated subset of the AIDev dataset~\cite{li2025aidev}, we compute review latency (in hours) as the difference between the PR creation timestamp and its closure timestamp. We exclude PRs with missing timestamps or non-positive durations to ensure reliable measurement. This analysis is conducted on two disjoint sets of Agentic-PRs: \textbf{1,130 closed PRs identified as security-related} using our keyword-based filtering and manual validation procedure, and \textbf{30,154 remaining closed Agentic-PRs} in the curated subset that were not identified as security-related by this procedure. 

To assess whether observed differences in review latency are statistically significant, we apply the non-parametric Mann--Whitney U test \cite{mann1947test}. This choice is motivated by the highly skewed distribution of PR review times and avoids assumptions of normality. We report in \secref{subsec:rq4_results}  descriptive statistics alongside statistical test results to support robust interpretation.

To account for variation in reviewer trust across agents, we further compute agent-specific median review latencies for security and non-security PRs. By comparing within-agent differences, we assess whether certain agents experience disproportionately higher or lower scrutiny when submitting security-relevant changes relative to their typical contributions.

To examine whether the scrutiny patterns observed for agent-authored security PRs differ from those for non-agent contributions, we also analyze human-authored PRs from the same repositories. The AIDev dataset provides PRs authored by human developers within the same GitHub repositories as the agent-authored PRs. Thus, we filter them to the 198 repositories that overlap with our agent security PR corpus, yielding 4,306 human-authored PRs. We apply the same keyword-based filtering described in Section \ref{subsec:rq1_method} to classify these PRs as security-relevant (841 PRs) or non-security (3,465 PRs). Then, we manually analyze them. This way, we have \textbf{634} human-authored security-relevant PRs. We compute merge rates and review latency for both human groups and compare them against the corresponding agent groups using Mann–Whitney U tests for latency and chi-squared tests for merge rates.

\subsection{Answering RQ5: Early Predictors of Risky or Rejected PRs}
\label{subsec:rq5_method}

To address \textbf{RQ5}, we investigate whether early, observable signals available at PR creation time can help predict whether a security-relevant Agentic-PR will ultimately be rejected.

\paragraph{Dataset and Target Definition}
We restrict this analysis to \textit{closed} security-relevant Agentic-PRs, since open PRs lack definitive outcomes. In our dataset of \textbf{1,293 security-relevant PRs},  \textbf{1,130} of them were closed. Following our dataset schema, we label a PR as \textit{rejected} if it is closed without being merged (\ie~\texttt{merged\_at} is missing), and as \textit{merged} otherwise. In this way, we have \textbf{435 rejected PRs} (38.5\%) and \textbf{695 merged PRs}. This binary outcome serves as a conservative proxy for perceived risk, insufficient confidence, or lack of acceptance of the agent-generated security contribution.

\paragraph{Early Feature Extraction (Structured Signals)}
From each PR, we extract lightweight structured signals that are available from the PR title and description at creation time, without relying on code churn or file-level metadata. Since AIDev does not uniformly provide additions, deletions, or file-touch counts for all PRs, we do not use code-churn-based size features. Instead, we replace the original \texttt{size} variable with five text-structure features derived from the PR description and title: body word count, number of Markdown sections (\texttt{\#\#}), title word count, number of code blocks (\texttt{```}), and number of task checkboxes. These features capture complementary aspects of PR scope, documentation detail, and implementation complexity as reflected in the PR text.

We further include two security-content signals: a binary indicator for whether the PR title contains a security keyword as defined in \secref{subsec:rq2_method} (\eg~\textit{auth}, \textit{crypto}, \textit{token}, \textit{password}, \textit{key}, \textit{credential}, \textit{payment}), and the count of security keywords appearing in the PR body. In addition, we encode two high-level security category indicators, \textbf{Vulnerability Fix} and \textbf{Dependency Update}, along with agent identity. Before training logistic regression models, all continuous features are standardized to reduce sensitivity to scale differences and outliers. We also conduct descriptive analysis by comparing feature distributions across merged and rejected PRs and examining the association between individual early signals and rejection outcomes.

\paragraph{Train/Test Split and Class Imbalance}
We use a stratified random split, holding out 30\% of PRs for testing and using the remaining 70\% for training. Because merged and rejected PRs are typically imbalanced, we apply class-balanced weighting during model training across all classifiers.

\paragraph{Structured-Feature Prediction Models}
We evaluate two structured-data classifiers. First, we train a \textbf{logistic regression} model with feature standardization as a simple, interpretable baseline~\cite{hosmer2013applied}. Second, we train a \textbf{Random Forest} classifier~\cite{breiman2001random} to capture potential non-linear relationships and feature interactions among early signals. We choose Random Forests because they (i) perform robustly on mixed-scale tabular features with minimal tuning, (ii) naturally model non-linearities and interactions, and (iii) provide feature-importance estimates that support interpretability. We report precision, recall, and F1-score on the held-out test set, and we analyze coefficient magnitudes (logistic regression) and feature importance (Random Forest) to identify the most influential early predictors.

\paragraph{Text-Based Prediction Models}
To evaluate whether PR text alone can predict rejection, we train three text classifiers using PR titles and descriptions concatenated into a single document. As a lightweight linear baseline, we train a \textbf{TF--IDF + linear SVM} classifier~\cite{salton1988term,joachims1998text} with class-balanced weighting. We then train a \textbf{fastText} supervised classifier~\cite{joulin2017fasttext} as an efficient embedding-based model suitable for short technical text. Finally, we fine-tune a pre-trained \textbf{DistilBERT} model~\cite{sanh2019distilbert} for binary sequence classification using standard training settings (max sequence length 128, 3 epochs, learning rate $2\times 10^{-5}$) and incorporate class weights in the loss function to account for imbalance. We compare performance across the structured-feature models and text-only models to assess whether semantic cues in PR text provide a predictive signal beyond lightweight metadata.

\paragraph{LLM-Based Prediction Models}
We further evaluate recent general-purpose LLMs as prompting-based baselines for security PR rejection prediction. Specifically, we use \textbf{GPT-4o-mini}~\cite{openai2024gpt4omini} in both zero-shot and few-shot settings as a larger-model comparison. Unlike the supervised models, these LLM baselines are not fine-tuned on the prediction task. Instead, each model receives only the PR title and description and is asked to predict whether maintainers will merge the PR or reject it by closing it without merging.

For the zero-shot setting, the system prompt describes the model as a GitHub pull-request reviewer assistant and instructs it to return a single JSON object with one of two labels: \texttt{\{"prediction": "merged"\}} or \texttt{\{"prediction": "rejected"\}}. The user prompt contains the PR title and the PR description. To control token cost and reduce prompt-length variation, descriptions are truncated to the first 300 words. For the few-shot setting, we prepend four examples sampled from the training split, with two merged and two rejected PRs. Each example includes the title, a truncated description of up to 100 words, and the corresponding JSON label. The model then predicts the label for a held-out test PR using the same output format. \section{Results}
\label{sec:results}
In this section, we present the answers to our research questions. 

\subsection{RQ1: Prevalence and Distribution of Security-Relevant Agentic PRs}
\label{subsec:rq1_results}

\begin{figure}[!htbp]
    \centering
    \includegraphics[width=\linewidth]{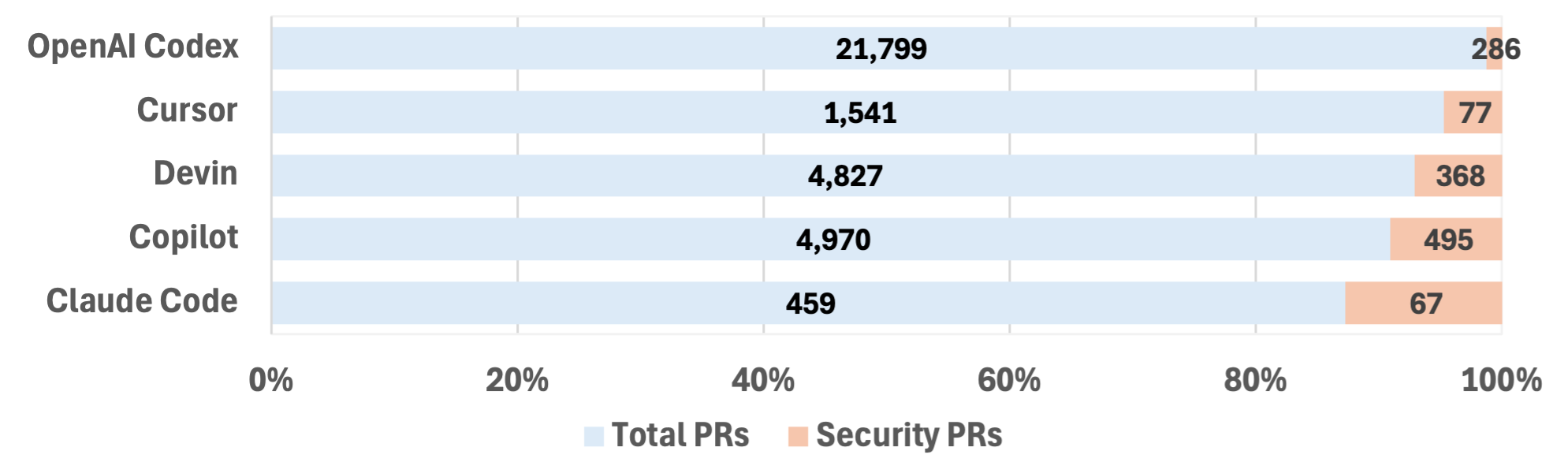}
    \caption{Prevalence of Security-relevant PRs Across Different Autonomous Coding Agents}
    \label{fig:rq1_results}
\end{figure}

We first examine how frequently autonomous coding agents contribute security-relevant pull requests and how these contributions are distributed across agents. 
Figure~\ref{fig:rq1_results} summarizes the prevalence of security-relevant Agentic-PRs. From the curated AIDev dataset, we identified \textbf{1,293 security-related PRs}. These 1,293 PRs represent \textbf{3.85\% of all 33,596 Agentic-PRs} in the curated dataset.

Security-relevant contributions are unevenly distributed across agents.  \textsf{Claude Code}  shows the highest proportion of security-relevant PRs (\textbf{459} PRs -- \textbf{14.6\%}), followed by \textsf{Copilot} (\textbf{4,970} PRs -- \textbf{10\%}) and \textsf{Devin} (\textbf{4,827} PRs -- \textbf{7.6\%}). \textsf{OpenAI Codex} is the agent with the smallest proportion of security-focused PRs, with only \textbf{1.3\%} of its PRs confirmed as security-relevant after manual validation.

\paragraph{Distribution Across Projects}
The dataset spans \textbf{2,807 unique repositories}, yet the distribution of Agentic-PRs is highly skewed (Gini coefficient = 0.83): just \textbf{22 repositories} account for 50\% of all PRs, and 306 repositories account for 80\%. The median repository contributes only \textbf{2 PRs}, while the most active single repository (\texttt{mochilang/mochi}) contributes 8,913. Security PRs are distributed more evenly across \textbf{520 repositories}, with 61.7\% of those repositories contributing a single security PR, and only 23 contributing ten or more.

This concentration is agent-specific: \textsf{OpenAI Codex} covers the broadest set of repositories (1,248 unique repos, avg.\ 17.5 PRs/repo), reflecting its pipeline-style bulk operation, while \textsf{Claude Code} shows the highest repository diversity relative to its volume (213 repos, avg.\ 2.2 PRs/repo), suggesting more targeted, one-off engagements.

\begin{figure}[!htbp]
    \centering
    \includegraphics[width=\linewidth]{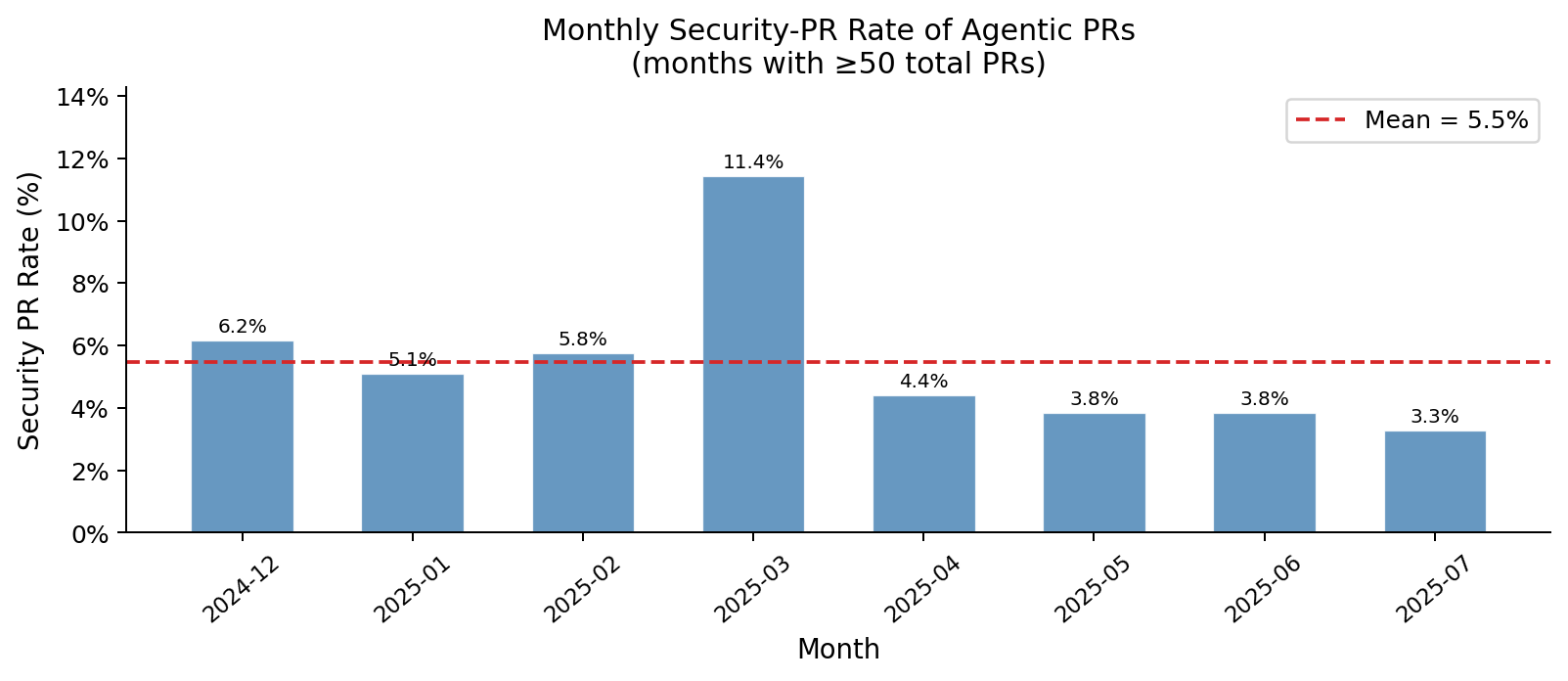}
    \caption{Monthly Security PR Rate of Agentic  PRs. }
    \label{fig:longitudinal}
\end{figure}

\paragraph{Longitudinal Stability}
The dataset spans \textbf{2024-Q4 through 2025-Q3}, with the large majority of PRs concentrated in Q2--Q3 2025 as adoption accelerated (Q1: 1,695 PRs; Q2: 18,697; Q3: 13,139).
Figure~\ref{fig:longitudinal} plots the monthly security-PR rate for confirmed security PRs.

The security-PR rate is broadly stable at \textbf{3.8--5.8\%} across most months, with an isolated spike to 11.4\% in March 2025 driven by a concentrated wave of \textsf{Devin} activity on security-sensitive repositories. Excluding that month, the rate remains in the 3.8--6.2\% band, suggesting that agents' propensity to generate security-relevant PRs has not materially changed as adoption grew.

\subsection{RQ2: Security Outcomes Across Agents, Ecosystems, and Code Types}
\label{subsec:rq2_results}

This RQ examines whether the merge rates of security-related PRs vary across autonomous coding agents, programming ecosystems, and types of code changes. We compute the merge rate as the \textit{number of merged PRs} divided by the \textit{total number of closed PRs} (merged or unmerged).

\begin{figure}
    \centering
    \includegraphics[width=\linewidth]{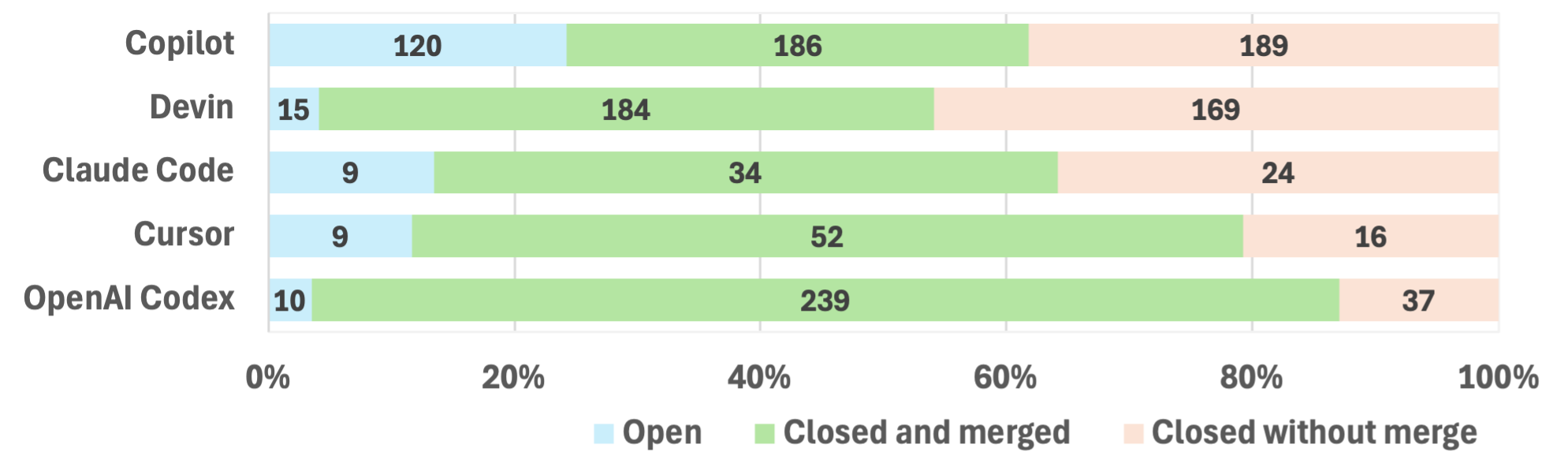}
    \caption{RQ2 Results -- Outcomes of Security-Related Pull Requests Authored by Agents.}
    \label{fig:rq2_agent_outcomes}
\end{figure}

\paragraph{Merge Rates Per Agent}
Figure~\ref{fig:rq2_agent_outcomes} summarizes the statuses of the security-related pull requests authored by each agent. 
As observed in this figure, we found that \textsf{OpenAI Codex} achieves the highest merge rate (86.59\%) followed by \textsf{Cursor} (76.47\%), \textsf{Claude Code} (58.62\%), and \textsf{Devin} (52.12\%) while \textsf{Copilot} (49.60\%) demonstrates the lowest merge rate.

\paragraph{Merge Rates Across Programming Languages}
Table~\ref{tab:rq2_language_outcomes} reports the most frequent programming languages among security-related Agentic-PRs and their corresponding merge rates, sorted by merge rate.
While \textsf{TypeScript} accounts for the largest number of security PRs (447 PRs), it exhibits a comparatively lower merge rate (56.51\%), suggesting stricter review or higher rejection likelihood in this ecosystem.
In contrast, \textsf{Python} security PRs (252 PRs) achieve a higher merge rate (68.30\%), indicating more favorable acceptance outcomes.

Languages with smaller numbers, such as \textsf{Ruby} and \textsf{HTML}, exhibit the highest merge rates (above 80\%); however, these results should be interpreted cautiously due to limited sample sizes. Mid-volume systems-oriented languages, including \textsf{Go}, \textsf{Java}, and \textsf{C\#}, show relatively consistent merge rates around 60--62\%, suggesting more uniform reviewer expectations in these ecosystems. Notably, \textsf{Rust} security PRs have the lowest merge rate (51.16\%), which may reflect heightened scrutiny in safety-critical or performance-sensitive contexts.
\begin{table}[!htbp]
\centering
\scriptsize
\caption{Top 10 languages for security PRs and their merge rates.}
\label{tab:rq2_language_outcomes}
\begin{tabular}{lcc}
\toprule
\textbf{Language} & \textbf{Total \# PRs} & \textbf{Merge Rate (\%)} \\
\midrule
Ruby        & 20  & 83.33 \\
HTML        & 26  & 82.61 \\
Python      & 252 & 68.30 \\
JavaScript  & 48  & 65.00 \\
Dart        & 19  & 63.16 \\
Go          & 108 & 61.70 \\
Java        & 48  & 60.00 \\
C\#         & 103 & 60.53 \\
TypeScript  & 447 & 56.51 \\
Rust        & 51  & 51.16 \\
\bottomrule
\end{tabular}
\end{table}

\paragraph{Merge Rates Per Domain}
Table~\ref{tab:rq2_domain_outcomes} aggregates security-related PRs into broader ecosystem domains. At the domain level, Data/ML exhibits the highest merge rate among the major domains (63.71\%), followed closely by Web (62.11\%) and Enterprise (60.18\%), while Systems has the lowest merge rate (56.15\%). Operations security PRs achieve the highest overall merge rate (69.70\%); however, it is important to highlight that this result is based on a very small number of PRs (35) and should therefore be interpreted with caution.

\begin{table}[!htbp]
\centering
\scriptsize
\caption{Security PR outcomes aggregated by ecosystem domain.}
\label{tab:rq2_domain_outcomes}
\setlength{\tabcolsep}{6pt}
\begin{tabular}{lcc}
\toprule
 \textbf{Domain} & \textbf{Total \# PRs} & \textbf{Merge Rate (\%)} \\
\midrule

  Web        & 606 & 62.11 \\
  Data/ML    & 262 & 63.71 \\
  Systems    & 216 & 56.15 \\
  Enterprise & 156 & 60.18 \\
  Operations        & 35   & 69.70 \\
\bottomrule
\end{tabular}
\end{table}

\paragraph{Distribution and Merge Rates of Security Change Types}
Table~\ref{tab:rq2_nature_by_agent} summarizes how autonomous coding agents distribute their security-related work across major security change categories and also reports aggregate category totals and merge rates. Overall, \emph{Security Feature} and \emph{Vulnerability Fix} PRs account for the largest share of security-related changes, with \emph{Security Feature} PRs being both the most prevalent and exhibiting higher aggregate merge rates than \emph{Vulnerability Fix} PRs. In contrast, \emph{Configuration/Compliance} and \emph{Dependency Update} PRs occur less frequently but achieve comparatively high merge rates.

At the agent level, variation emerges in how security work is distributed. For example, \textsf{Claude Code} predominantly contributes \emph{Security Feature} PRs (68.7\%), whereas \textsf{Copilot} places greater emphasis on \emph{Vulnerability Fix} PRs (50.5\%). \textsf{Cursor} and \textsf{OpenAI Codex} exhibit more balanced distributions between these two categories while also contributing a non-trivial share of \emph{Configuration/Compliance} PRs. Overall, these patterns indicate that agents prioritize distinct forms of security work rather than converging on a uniform security contribution profile.
\begin{table}[!htbp]
\centering
\scriptsize
\caption{Nature of security work by agent: percentage breakdown of categories within each agent's security PRs. Aggregate category volumes and merge rates are shown at the top.}
\label{tab:rq2_nature_by_agent}
\setlength{\tabcolsep}{5pt}
\begin{tabular}{lcccc}
\toprule
\textbf{Agent / Aggregate} 
& \textbf{Sec. Feat.} 
& \textbf{Vuln. Fix} 
& \textbf{Conf./Compl.} 
& \textbf{Depn. Update} \\
\midrule
\textbf{Total \# PRs} 
& 554 & 506 & 134 & 113 \\

\textbf{Merge Rate (\%)} 
& 64.61 & 53.94 & 70.08 & 65.66 \\
\midrule
Claude Code   & 68.7 & 25.4 & 0.0  & 6.0 \\
Copilot       & 33.1 & 50.5 & 3.8  & 12.5 \\
Cursor        & 40.3 & 41.6 & 14.3 & 3.9 \\
Devin         & 51.6 & 27.5 & 13.9 & 7.1 \\
OpenAI Codex  & 40.6 & 35.7 & 17.8 & 5.9 \\
\bottomrule
\end{tabular}
\end{table}
 
\paragraph{Agent-level variation}
Table~\ref{tab:agent_profile} shows that security engagement varies widely across agents:
security PRs account for only 1.3\% of OpenAI Codex's closed PRs but 14.6\% of
Claude Code's. Three patterns help explain this variation.

\begin{table*}[t]
\centering
\scriptsize
\setlength{\tabcolsep}{2.5pt}
\caption{Per-agent characteristics for security PRs, ordered by security PR rate (descending). Security rate = confirmed security PRs / all closed PRs for that agent. Issue-routed = security PRs linked to a pre-existing issue. Domain percentages are computed over confirmed security PRs. Gap = non-security minus security merge rate (pp).}
\label{tab:agent_profile}
\begin{tabular}{lccccccl}
\toprule
\textbf{Agent} & \textbf{PRs} & \textbf{Sec. \%} &
  \multicolumn{2}{c}{\textbf{Merge \%}} & \textbf{Gap} &
  \textbf{Issue \%} & \textbf{Top domain} \\
\cmidrule(lr){4-5}
 & & & Sec & Non & & & \\
\midrule
Claude Code  &   459    & 14.6 & 58.6 & 73.6 & $+$15.0 & 17.9 & Web (43.3)     \\
Copilot      & 4{,}970  & 10.0 & 49.6 & 55.5 & $+$5.9  & 81.2 & Web (34.7)     \\
Devin        & 4{,}827  &  7.6 & 52.1 & 55.8 & $+$3.7  &  6.8 & Web (75.8)     \\
Cursor       & 1{,}541  &  5.0 & 76.5 & 74.5 & $-$2.0  &  9.1 & Web (61.0)     \\
Codex        & 21{,}799 &  1.3 & 86.6 & 85.8 & $-$0.8  &  2.4 & Data/ML (38.1) \\
\bottomrule
\end{tabular}
\end{table*}
 
\textbf{Integration mode.}
Table~\ref{tab:agent_profile} shows that Copilot is the most issue-driven agent: 81.2\% of its security PRs are linked to a pre-existing issue. This suggests that Copilot often responds to human-scoped security concerns rather than selecting security tasks independently. In contrast, OpenAI Codex and Cursor have much lower issue-routing rates, 2.4\% and 9.1\%, respectively, indicating a more self-directed mode of operation. This difference is also reflected in outcomes: Codex has the highest security merge rate (86.6\%) and almost no security penalty relative to non-security PRs ($-0.8$ pp), consistent with narrower, more selective fixes.

\textbf{Project domain.}
The dominant domain also differs by agent. Codex's security PRs are most often in Data/ML repositories (38.1\%), whereas Devin, Cursor, Claude Code, and Copilot are led by Web projects, with Web accounting for 75.8\%, 61.0\%, 43.3\%, and 34.7\% of their security PRs, respectively. This distribution suggests that agents are not engaging with the same security workload: some operate mainly in Web settings, where security fixes may involve authentication, input handling, or request-flow context, while Codex's activity is more concentrated in Data/ML repositories.

\textbf{Security breadth versus acceptance.}
The security rate and merge rate capture different aspects of agent behavior. Claude Code has the highest security-PR rate (14.6\%) but also the largest merge-rate gap: its security PRs merge at 58.6\%, compared with 73.6\% for non-security PRs ($+15.0$ pp). This suggests broader security engagement but lower acceptance once those changes reach review. Codex shows the opposite pattern: it has the lowest security-PR rate (1.3\%) but the highest security merge rate (86.6\%). Thus, security engagement and acceptance are not interchangeable measures. The former reflects how often an agent attempts security work, whereas the latter reflects how readily those attempts are accepted by maintainers.

\paragraph{Longitudinal Stability for Merge Rate}
The quarterly merge rate for security PRs rises from 42.2\% in Q1 2025 to \textbf{66.3\%} in Q2 and stabilizes at 60.8\% in Q3, consistent with two confounding factors: (i)~\textsf{Devin} dominates Q1 (133 of 135 closed security PRs) and exhibits one of the lower per-agent merge rates (42.4\% in Q1), naturally depressing the aggregate; and (ii)~Q2 onward introduces \textsf{Copilot} and \textsf{OpenAI Codex} at scale, both with distinct merge dynamics (Codex: 84.7\% $\to$ 92.5\%; Copilot: 48.5\% $\to$ 50.9\%).

\subsection{RQ3 Results: Security Actions and Intents in Agentic-PRs}
\label{subsec:rq3_results}

After open coding 1,293 manually confirmed security-relevant Agentic-PRs from the curated AIDev subset, we identify recurring \emph{security actions} (what agents do) and \emph{security intents} (why agents do it). We report only themes that appear in at least 10 PRs to focus on stable and recurrent patterns.

\subsubsection{Security Actions Performed by Autonomous Agents}
\label{subsec:rq3_actions}

\paragraph{Distribution of Security Actions and Intents}
Table~\ref{tab:rq3_actions_intents} summarizes the most frequent \emph{security actions} and \emph{security intents} identified through open coding across the 1,293 security-relevant Agentic-PRs. Actions capture \emph{how} agents modify the codebase, while intents capture \emph{why} those changes are made.

\begin{table}[!htbp]
\centering
\scriptsize
\caption{Security actions and intents identified via open coding (themes with fewer than 10 occurrences omitted).}
\label{tab:rq3_actions_intents}
\setlength{\tabcolsep}{4pt}
\begin{tabular}{l c l c}
\toprule
\multicolumn{4}{c}{\textbf{Actions}} \\
\cmidrule(lr){1-4}
\textbf{Theme} & \textbf{\#} &
\textbf{Theme} & \textbf{\#} \\
\midrule
Code Refactoring & 957 &
Testing & 755 \\

Documentation & 692 &
Error Handling & 595 \\

Security Improvements & 580 &
API Development & 540 \\

Configuration Management & 495 &
User Experience Enhancement & 484 \\

Input Validation & 434 &
Authentication \& Authorization & 377 \\

Dependency Management & 367 &
Performance Optimization & 313 \\

Logging and Monitoring & 262 &
Compatibility Maintenance & 179 \\

Version Control & 143 &
Database Management & 49 \\

Others & 28 &
 &  \\
\midrule
\multicolumn{4}{c}{\textbf{Intents}} \\
\cmidrule(lr){1-4}
\textbf{Theme} & \textbf{\#} &
\textbf{Theme} & \textbf{\#} \\
\midrule
Functionality Improvement & 890 &
Vulnerability Mitigation & 741 \\

User Experience & 697 &
Error Handling & 604 \\

Security Enhancement & 531 &
Testing and Reliability & 468 \\

Compatibility Assurance & 457 &
Performance Optimization & 389 \\

Code Quality & 376 &
User Guidance & 340 \\

Maintainability & 335 &
Compliance and Standards & 255 \\

Documentation Improvement & 218 &
Development Efficiency & 183 \\

Resource Management & 82 \\
\bottomrule
\end{tabular}
\end{table}
 Overall, agentic security work is dominated by a small number of recurring implementation patterns. On the action side, \emph{Code Refactoring} (957), \emph{Testing} (755), and \emph{Documentation} (692) are most prevalent, indicating that security improvements are often realized through restructuring existing code, adding or extending tests, and clarifying usage or constraints rather than introducing entirely new mechanisms. Lower-level but still common actions such as \emph{Error Handling}, \emph{Configuration Management}, and \emph{Input Validation} further suggest a focus on robustness and defensive hardening.

\begin{figure}[!htbp]
    \centering
    \includegraphics[width=0.85\linewidth]{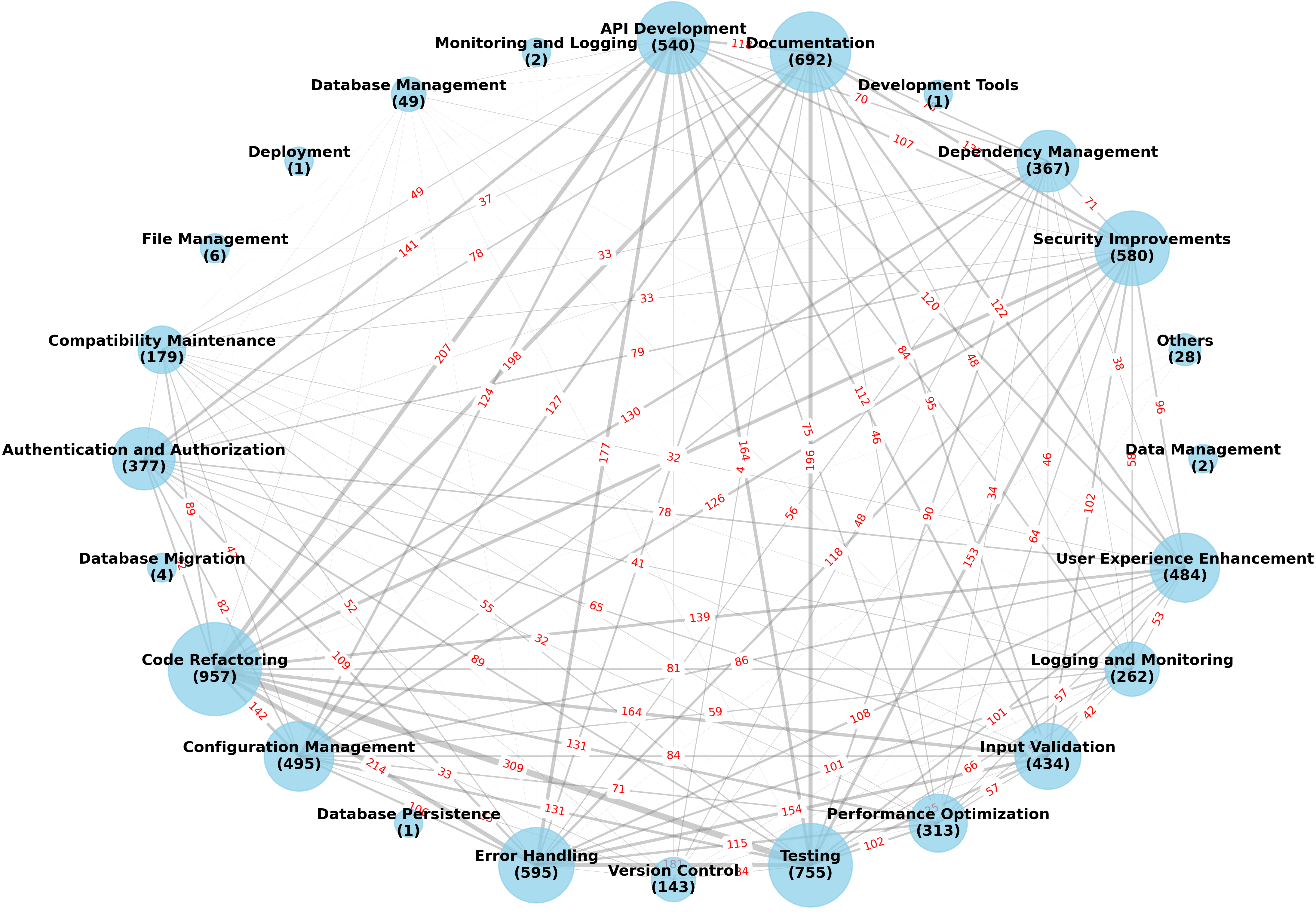}
    \caption{Co-occurrence Network of Security Actions. }
    \label{fig:action}
\end{figure}

On the intent side, \emph{Functionality Improvement} (890) and \emph{Vulnerability Mitigation} (741) are the two dominant motivations. Notably, a large fraction of security-relevant PRs are framed as improving functionality or developer experience rather than narrowly fixing a specific vulnerability, reinforcing that security work in practice is often intertwined with broader maintenance and evolution goals. Intents related to \emph{User Experience}, \emph{Reliability}, and \emph{Compatibility Assurance} also appear frequently, reflecting a preventive and quality-oriented framing of security changes.

\begin{figure}[!htbp]
    \centering
    \includegraphics[width=0.85\linewidth]{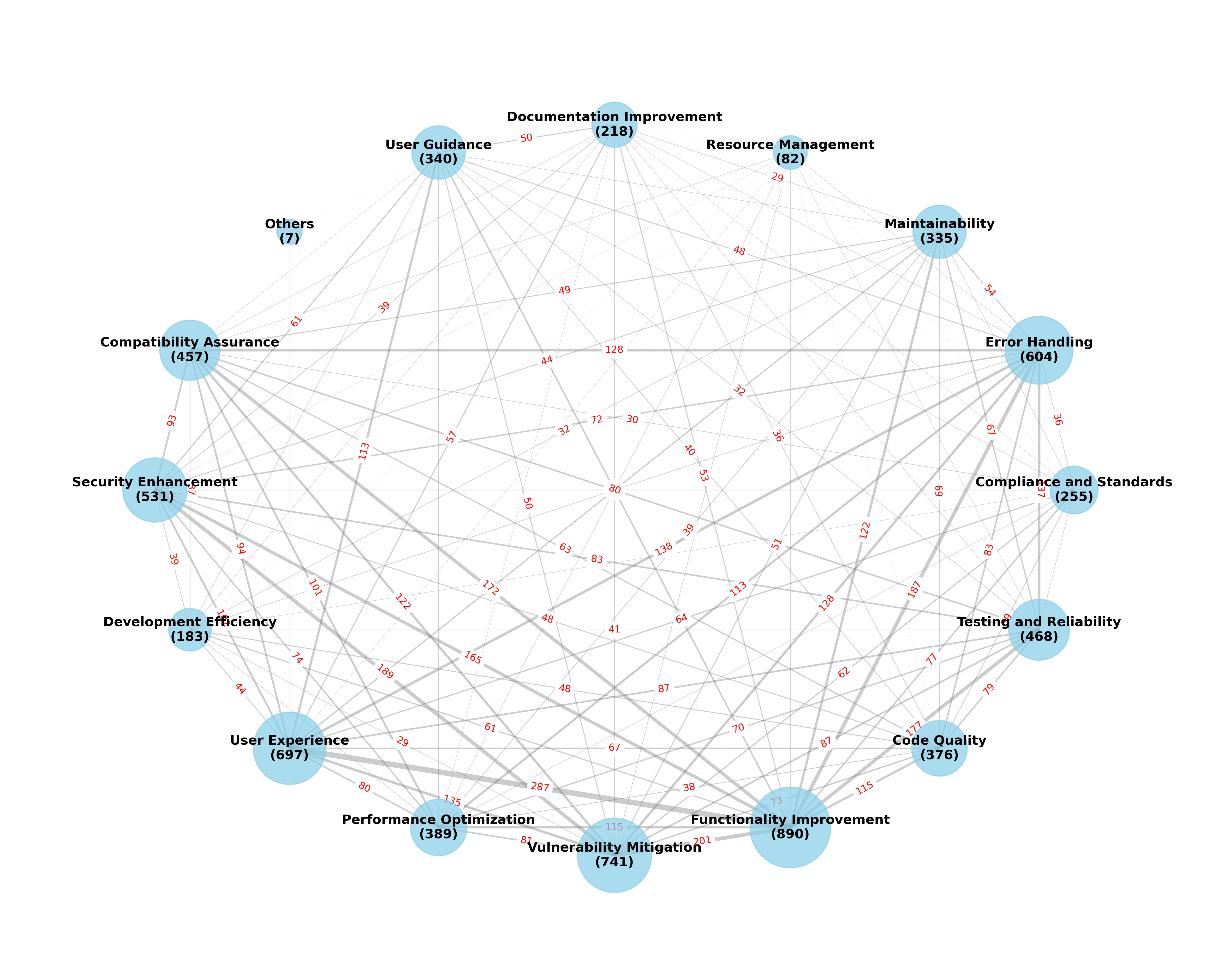}
    \caption{Co-occurrence Network of Security Intents. }
    \label{fig:intent}
\end{figure}

\paragraph{Action–Intent Co-occurrence Patterns}
To move beyond marginal frequencies, we analyze how security-related actions and intents co-occur within the same Agentic-PRs. We quantify these associations using pairwise \emph{lift}, which measures how much more frequently two themes co-occur than would be expected if they were independent, and the $\phi$ coefficient, which captures the strength of correlation between two binary-coded themes. We identify recurring and statistically strong links between \emph{how} agents modify code and \emph{why} those changes are made. The resulting co-occurrence patterns show that agentic security work is rarely one-dimensional. Instead, agents systematically pair concrete implementation techniques with aligned security motivations. In Figure \ref{fig:action} and Figure \ref{fig:intent}, we present their co-occurrence networks. To complement the co-occurrence network visualization and improve readability, Table~\ref{tab:rq3_cooccurrence_links} summarizes representative high-strength co-occurrence links identified in the RQ3 analysis.

\begin{table*}[!htbp]
\centering
\scriptsize
\caption{Strong co-occurrence links from the RQ3 action--intent analysis.}
\label{tab:rq3_cooccurrence_links}
\setlength{\tabcolsep}{5pt}
\begin{tabular}{llcc}
\toprule
\textbf{Theme 1} & \textbf{Theme 2} & \textbf{Lift} & \textbf{$\phi$} \\
\midrule
Documentation & User Guidance & 2.77 & 0.41 \\
Documentation & Documentation Improvement & 2.70 & 0.31 \\
User Experience Enhancement & User Experience & 2.50 & 0.43 \\
Performance Optimization & Performance Optimization & 3.50 & 0.39 \\
Performance Optimization & Resource Management & 3.54 & 0.17 \\
Error Handling & Error Handling & 2.05 & 0.31 \\
Dependency Management & Version Control & 2.38 & -- \\
API Development & Database Management & 2.46 & -- \\
\bottomrule
\end{tabular}
\end{table*}

From the results, we found several strong and intuitive associations. \emph{Documentation} actions exhibit one of the strongest links to intent, co-occurring frequently with both \emph{User Guidance} (lift = 2.77, $\phi$ = 0.41) and \emph{Documentation Improvement} (lift = 2.70, $\phi$ = 0.31). This indicates that documentation-related security PRs are commonly motivated by steering correct and secure usage rather than by purely descriptive goals. Similarly, \emph{User Experience Enhancement} actions show a strong association with \emph{User Experience} intents (lift = 2.50, $\phi$ = 0.43), suggesting that agents often frame security improvements as usability fixes that reduce misuse or error-prone interactions.

Performance-related work displays particularly strong coupling. \emph{Performance Optimization} actions co-occur most frequently with both \emph{Performance Optimization} intents (lift = 3.50, $\phi$ = 0.39) and \emph{Resource Management} intents (lift = 3.54, $\phi$ = 0.17). These high-lift associations suggest that performance tuning is often treated as security-relevant when it mitigates resource exhaustion, reliability issues, or denial-of-service risks. Likewise, \emph{Error Handling} actions strongly align with \emph{Error Handling} intents (lift = 2.05, $\phi$ = 0.31), reflecting direct remediation of failure modes that could otherwise be exploited.

Beyond direct action–intent pairs, we also observe meaningful co-occurrence among actions themselves. For example, \emph{Dependency Management} frequently co-occurs with \emph{Version Control} actions (lift = 2.38), and \emph{API Development} co-occurs with \emph{Database Management} (lift = 2.46), indicating that security-related changes often span multiple technical layers within a single PR.
Overall, these co-occurrence patterns demonstrate that agentic security PRs are not collections of isolated fixes. Instead, autonomous agents consistently align specific implementation choices with coherent security rationales. This reinforces the analytical value of separating \emph{actions} from \emph{intents} and shows that agent-authored security work blends vulnerability remediation, preventive hardening, and quality improvement within broader software engineering activities.

\paragraph{Intent--Action Alignment}
To assess whether intent and action labels co-occur coherently, we classify each PR into one of four alignment quadrants based on whether its open-coding labels include at least one security-oriented intent (\textit{Security Enhancement}, \textit{Vulnerability Mitigation}, \textit{Compliance and Standards}) and at least one security-oriented action (\textit{Security Improvements}, \textit{Authentication and Authorization}, \textit{Input Validation}, \textit{Logging and Monitoring}).

Notably, \textbf{intent--action misalignment does not predict rejection}. Merge rates across the four quadrants are nearly identical: fully aligned PRs (security intent + security action, $n=743$) merge at \textbf{62.6\%}, PRs with a security intent but general action ($n=342$) at \textbf{63.5\%}, PRs with a general intent but security action ($n=388$) at \textbf{62.6\%}, and PRs with neither ($n=777$) at \textbf{62.5\%}. This uniformity suggests that reviewers respond to the \emph{presence} of security-sensitive code changes, not to whether the agent's stated motivation is security-framed.

The most common misalignment pattern is \textit{Authentication and Authorization} actions paired with \textit{Functionality Improvement} intent: agents frequently introduce or extend authentication mechanisms as part of a broader feature addition, without framing the PR as a security change. Similarly, \textit{Input Validation} frequently appears alongside \textit{Development Efficiency} or \textit{Code Quality} intents, reflecting security hardening bundled into general refactoring work. These cases illustrate that agent-authored security work is often \emph{embedded} in multi-purpose PRs rather than isolated as dedicated security fixes.

\paragraph{Non-Security Intents in Security PRs}
Two intents, \textit{Performance Optimization} (1,242 co-occurrences) and \textit{Development Efficiency} (663), appear frequently despite seeming only tangentially related to security. Both arise because agent-authored security PRs are rarely single-purpose: the most common actions co-occurring with \textit{Performance Optimization} intent are \textit{Code Refactoring} (13\%), \textit{Error Handling} (10\%), and \textit{Testing} (9\%), with \textit{Security Improvements} accounting for 7\%. Among confirmed security PRs specifically, \textit{Performance Optimization} appears in 157 cases, primarily when fixing a vulnerable code path simultaneously removes an inefficiency (\eg~replacing a vulnerable regex with a hardened, faster implementation). \textit{Development Efficiency} co-occurs most with \textit{Documentation} (14\%) and \textit{Testing} (11\%), reflecting agents that add security-related tests or update documentation while addressing a vulnerability. These intents therefore capture real bundling behavior in agent-authored PRs: agents embed security changes within broader improvement tasks rather than submitting focused, single-intent patches.

\subsection{RQ4: Reviewer Behavior Toward Security Agentic-PRs}
\label{subsec:rq4_results}

To understand whether human reviewers apply different levels of scrutiny to security-relevant Agentic-PRs compared to non-security Agentic-PRs. We report results along three dimensions: (i) aggregate review outcomes and latency, (ii) interaction-based scrutiny signals, and (iii) agent- and category-level heterogeneity.

\paragraph{Aggregate Review Outcomes and Latency}

Table~\ref{tab:rq4_latency_overall} summarizes review latency statistics. From the results, we found that security-relevant Agentic-PRs have consistently lower merge rates than non-security PRs; \ie~security PRs are merged at only \textbf{61.50\%}, compared to \textbf{77.33\%} for non-security PRs. We also found that Security PRs have a median review latency of \textbf{3.92} hours, compared to only \textbf{0.11} hours for non-security PRs, and a mean latency more than twice as long (97.45 vs.\ 38.29 hours). This difference is statistically significant ($p < 0.001$), indicating substantially increased review time for confirmed security-relevant Agentic-PRs.
\begin{table}[!htbp]
\centering
\scriptsize
\caption{Review latency (hours) for security vs.\ non-security merged Agentic-PRs.}
\label{tab:rq4_latency_overall}
\setlength{\tabcolsep}{6pt}
\begin{tabular}{lcccc}
\toprule
 \textbf{PR Type} & \textbf{Total \# PRs} & \textbf{Mean} & \textbf{Median} & \textbf{Std.\ Dev.} \\
\midrule
 Security     & 1,130  & 97.45 & 3.92 & 239.91 \\
  Non-Security & 30,154 & 38.29 & 0.11 & 140.42 \\
\bottomrule
\end{tabular}
\end{table} 
\paragraph{Interaction-Based Scrutiny Signals}

Beyond time and acceptance outcomes, we examine direct signals of reviewer engagement.
Security Agentic-PRs receive \textbf{more than twice as many review comments} on average (2.5 vs.\ 1.0) and trigger \textbf{over twice as many formal review events} (1.8 vs.\ 0.8), with statistically significant differences across all interaction types ($p < 0.001$).
Notably, security PRs are more likely to receive explicit \emph{Changes Requested} actions (0.1 vs.\ 0.0) and approvals accompanied by discussion, indicating active reviewer intervention rather than passive acknowledgment. Together, these interaction patterns indicate that increased latency is not merely idle waiting time, but corresponds to deeper reviewer engagement, discussion, and corrective feedback when agents modify security-sensitive code.

\paragraph{Agent-Level Heterogeneity}

Agent-level analysis (Table~\ref{tab:rq4_agent_latency}) shows that this effect holds across all agents, though with heterogeneous magnitude. Security-related PRs authored by \textsf{Devin} experience the largest median delay (+16.08 hours), followed by \textsf{Copilot} (+5.46 hours) and \textsf{Claude Code} (+3.42 hours), while \textsf{OpenAI Codex} exhibits the smallest observed median latency difference in our dataset.

To better understand this unusually low aggregate latency for Codex, we further examine the role of PR scope using the size-stratified analysis and agent-level description-length comparisons. These analyses indicate that review outcomes are strongly associated with PR size, and that Codex security PRs in our dataset tend to have  shorter descriptions than those of other agents, suggesting a narrower PR scope. Accordingly, the aggregate Codex latency should be interpreted cautiously as a property of the observed Codex security PR corpus in our dataset rather than as evidence that Codex-authored security PRs are inherently reviewed faster across projects.

\begin{table}[!htbp]
\centering
\scriptsize
\caption{Agent-specific review latency (hours) for security vs.\ non-security PRs.}
\label{tab:rq4_agent_latency}
\setlength{\tabcolsep}{4pt}
\begin{tabular}{lccccc}
\toprule
\textbf{Agent} & \textbf{Sec. Med.} & \textbf{Sec. Std.} & \textbf{Non-Sec. Med.} & \textbf{Non-Sec. Std.} & \textbf{Delta} \\
\midrule
Devin        & 23.76 & 312.60 & 7.68  & 173.18 & +16.08 \\
Copilot      & 18.32 & 241.49 & 12.87 & 215.95 & +5.46  \\
Claude Code  & 5.13  & 139.22 & 1.71  & 213.15 & +3.42  \\
Cursor       & 1.08  & 93.23  & 0.89  & 95.41  & +0.18  \\
OpenAI Codex & 0.07  & 125.87 & 0.02  & 109.16 & +0.05  \\
\bottomrule
\end{tabular}
\end{table} Merge rates further reinforce this pattern. While Codex security PRs are merged at a high rate (86.6\%), Copilot and Devin security PRs are merged in only about half of cases.
These differences suggest that reviewer scrutiny is shaped not only by security relevance but also by prior experience and perceived reliability of specific agents.

\paragraph{Security Action Categories}
To move beyond aggregate counts, we further examine how review latency varies across different \emph{security action categories}.
Table~\ref{tab:rq4_latency_by_category} summarizes review latency statistics conditioned on the primary type of security change.

\begin{table}[!htbp]
\centering
\scriptsize
\caption{Review latency (hours) for security-related Agentic-PRs by security change category.}
\label{tab:rq4_latency_by_category}
\setlength{\tabcolsep}{6pt}
\begin{tabular}{lcccc}
\toprule
\textbf{Category} & \textbf{Total \# PRs} & \textbf{Mean} & \textbf{Std.\ Dev.} & \textbf{Median} \\
\midrule
Config / Compliance & 127 & 127.17 & 341.67 & 0.86 \\
Security Feature    & 486 & 105.53 & 253.86 & 9.95 \\
Vulnerability Fix   & 432 & 84.19  & 190.43 & 3.35 \\
Dependency Update   & 99  & 83.40  & 185.83 & 4.36 \\
\bottomrule
\end{tabular}
\end{table}
 
Security Feature PRs exhibit the longest median review latency (\textbf{9.95} hours), substantially exceeding that of Vulnerability Fix PRs (\textbf{3.35} hours) and Dependency Update PRs (\textbf{4.36} hours).
In contrast, Config/Compliance PRs are reviewed markedly faster, with a median latency of only \textbf{0.86} hours.
These differences suggest that reviewers distinguish between classes of security work: changes that introduce or extend security mechanisms tend to trigger deeper inspection and longer deliberation, whereas configuration-level or compliance-oriented changes are often resolved more quickly.

\paragraph{Role of PR Complexity}

Finally, we assess whether heightened scrutiny can be explained solely by PR size or complexity.
Security Agentic-PRs are indeed larger and more complex, touching nearly \textbf{twice as many files} (30.3 vs.\ 15.4), exhibiting \textbf{higher code churn} (5,301 vs.\ 1,556 lines), and containing \textbf{longer descriptions} (256.6 vs.\ 97.2 words), with all differences statistically significant.

To assess whether heightened scrutiny persists \emph{after controlling for size}, we stratify all closed PRs into tertiles based on description length (small: $<$36 words; medium: 36--69 words; large: $>$69 words) and compare merge rates within each stratum (Table~\ref{tab:size_stratified}).

\begin{table}[t]
\scriptsize
\centering
\caption{Merge rate by PR size (description-length tertile) for closed security and
         non-security Agentic-PRs. Small: $<$36 words; Medium: 36--69 words;
         Large: $>$69 words. The aggregate security--non-security gap vanishes for
         small and medium PRs and is substantially reduced for large ones.}
\label{tab:size_stratified}
\begin{tabular}{lrrrrrr}
\toprule
\textbf{Size} & \multicolumn{2}{c}{\textbf{Security PRs}} & \multicolumn{2}{c}{\textbf{Non-Security PRs}} & \textbf{Gap} & \textbf{$p$} \\
\cmidrule(lr){2-3}\cmidrule(lr){4-5}
 & $n$ & MR (\%) & $n$ & MR (\%) & (pp) & \\
\midrule
Small  & 125 & 86.4 & 10{,}175 & 84.6 & $-$1.8 & 0.67 \\
Medium & 153 & 86.3 & 11{,}220 & 84.1 & $-$2.2 & 0.53 \\
Large  & 852 & 53.4 &  8{,}759 & 60.2 & $+$6.8 & $<$0.001$^{*}$ \\
\midrule
\textit{Overall} & \textit{1,130} & \textit{61.5} & \textit{30,154} & \textit{77.3} & \textit{$+$15.8} & \textit{$<$0.001}$^{*}$ \\
\bottomrule
\end{tabular}
\end{table}
 
The size-stratified analysis reveals that \textbf{the aggregate security--non-security merge gap is largely driven by PR size}. For small and medium PRs, security and non-security merge rates are statistically indistinguishable (small: 86.4\% vs.\ 84.6\%, $p=0.67$; medium: 86.3\% vs.\ 84.1\%, $p=0.53$). The gap emerges only among large PRs (53.4\% vs.\ 60.2\%, $p < 0.001$), where security PRs still merge \textbf{6.8 percentage points} less often, suggesting a residual scrutiny effect for complex security changes. However, because the majority of security PRs fall in the large stratum ($n=852$ of 1,130 closed), the aggregate 15.8 pp gap reflects primarily a size confound rather than a uniform security-content penalty: same-sized security and non-security PRs are treated with comparable acceptance rates. Time-to-close mirrors this pattern: among large PRs, security PRs take a median \textbf{16.5 hours} to close vs.\ \textbf{5.7 hours} for non-security PRs of equivalent size, while small and medium PRs close within minutes regardless of security relevance.

This complexity gap is further pronounced at the agent level. Using description length as a within-dataset proxy, we find a 10$\times$ spread: \textsf{OpenAI Codex} security PRs have a median description of 42 words, versus 336 words for \textsf{Copilot} and 421 words for \textsf{Claude Code}. Critically, this difference persists \emph{within each semantic category}: for Vulnerability Fix PRs alone, the median body length is 37 words for Codex versus 356 words for Copilot and 477 words for Claude Code, while Codex's merge rate (84\%) and median latency ($<$6 minutes) remain uniform across all four categories. These patterns suggest that the current four-category scheme captures \emph{intent} but not \emph{scope}: agents with lower contribution rates tend to handle simpler variants of each security type, and a finer-grained complexity dimension, such as lines changed or number of files affected, would be needed to isolate the contribution of reviewer scrutiny from task self-selection in future work.

However, the persistence of higher review latency, lower merge rates, and increased reviewer interaction, both across agents and within comparable security categories, suggests that reviewers respond not only to PR scale, but also to the perceived risk and responsibility associated with agent-authored security changes.

\paragraph{Comparing with Human-authored Security PRs}
We further compare security Agentic-PRs with \textbf{634} manually confirmed human-authored security PRs from the same repositories, and additionally examine how human reviewers treat security PRs relative to non-security PRs authored by humans. Among human-authored PRs, security PRs are accepted at a rate of \textbf{79.2\%}, compared to \textbf{83.5\%} for non-security PRs, a \textbf{4.3 percentage-point} gap that is statistically significant ($p = 0.011$). While human-authored security PRs also take longer to close (median \textbf{7.00} vs.\ \textbf{4.41} hours, $p < 0.001$), the absolute reduction in acceptance is modest.

This establishes an important baseline: security content causes a small but measurable dip in acceptance rates even for human authors, reflecting the inherently higher stakes of security-sensitive changes. In contrast, agentic-authored security PRs are merged at only \textbf{61.5\%}, a \textbf{17.7 percentage-point (pp)} gap relative to human security PRs ($p < 0.001$). The agent security penalty, measured as the gap between non-security and security merge rates within each author type, is \textbf{15.8 pp} for agents versus only \textbf{4.3 pp} for humans, making the agent-specific security scrutiny effect \textbf{3.7$\times$} larger. Furthermore, even for non-security PRs, agents are merged less often than humans (\textbf{77.3\%} vs.\ \textbf{83.5\%}, $\Delta = 6.2$ pp, $p < 0.001$), confirming a general reviewer wariness toward automated contributions. The amplification of this wariness for security content, from 6.2 pp to 17.7 pp, indicates that reviewers apply more heightened scrutiny to agent-authored security changes beyond any baseline distrust of automation.

\subsection{RQ5: Early Predictors of Risky or Rejected Security Agentic-PRs}
\label{subsec:rq5_results}

This research question examines whether early, observable signals available at pull-request creation time can help predict whether a security-relevant Agentic-PR will be rejected (\ie closed without merge). Our analysis focuses on the \emph{curated} dataset of manually validated security-related Agentic-PRs. In total, we analyze \textbf{1,130} closed security PRs, of which \textbf{435} are rejected, corresponding to a rejection rate of \textbf{38.5\%}.

\begin{table}[!htbp]
\centering
\scriptsize
\caption{Association between early signals and PR rejection in the curated security PR dataset (logistic regression coefficients).}
\label{tab:rq5_corr}
\setlength{\tabcolsep}{6pt}
\begin{tabular}{lr}
\toprule
\textbf{Feature} & \textbf{Coefficient ($\beta$)} \\
\midrule
Title length              & 0.316 \\
PR size                   & 0.049 \\
Description length        & 0.049 \\
Sensitive keyword in title & $-$0.054 \\
Agent (encoded)           & $-$0.268 \\
\bottomrule
\end{tabular}
\end{table}
 
Among the examined early signals, title length shows the strongest positive association with PR rejection, exhibiting the largest logistic regression coefficient ($\beta = 0.316$), followed by PR size and description length (both $\beta \approx 0.049$). Consistent with these associations, rejected PRs tend to be larger and more verbose on average: rejected PRs involve a mean of 2,720.3 lines changed compared to 1,873.1 for merged PRs, and have longer titles on average (59.8 vs.\ 52.7 characters). Together, these results suggest that early signals related to PR complexity and explanatory burden are associated with a higher likelihood of rejection.
In contrast, the presence of sensitive security-related keywords in PR titles exhibits a weak negative association with rejection ($\beta = -0.054$) and low importance across predictive models. This indicates that explicitly referencing security-sensitive topics (\eg~authentication, cryptography, credentials) is not, by itself, a strong early signal of rejection for security-related Agentic PRs.

\begin{table}[!htbp]
\centering
\scriptsize
\caption{Predictive performance for rejection prediction on the curated security-related
Agentic-PR dataset (held-out test set, $n=339$, 38.3\% rejected).
Metrics are reported for the \emph{Rejected} class.}
\label{tab:rq5_models}
\setlength{\tabcolsep}{6pt}
\begin{tabular}{lcccc}
\toprule
\textbf{Model}
& \textbf{Accuracy}
& \textbf{Rej. Precision}
& \textbf{Rej. Recall}
& \textbf{Rej. F1} \\
\midrule
\multicolumn{5}{l}{\textit{Structured-feature models}} \\
\quad Logistic Regression   & 0.58 & 0.46 & 0.48 & 0.47 \\
\quad Random Forest         & 0.61 & 0.49 & 0.41 & 0.45 \\
\midrule
\multicolumn{5}{l}{\textit{Text-based models (trained)}} \\
\quad SVM (TF--IDF)         & 0.65 & 0.53 & 0.72 & 0.61 \\
\quad fastText              & \textbf{0.71} & \textbf{0.62} & 0.60 & 0.61 \\
\quad DistilBERT (fine-tuned) & 0.64 & 0.52 & \textbf{0.85} & \textbf{0.64} \\
\midrule
\multicolumn{5}{l}{\textit{LLM baselines}} \\
\quad GPT-4o-mini (zero-shot) & 0.60 & 0.45 & 0.18 & 0.26 \\
\quad GPT-4o-mini (4-shot)    & 0.54 & 0.34 & 0.20 & 0.25 \\
\bottomrule
\end{tabular}
\end{table}
 We next evaluate the predictive power of early signals using structured-feature, text-based models, and prompting-based LLM baselines (Table~\ref{tab:rq5_models}). Among structured-feature approaches, logistic regression and Random Forest achieve modest performance, with accuracies of 0.58 and 0.61, respectively, and rejected-class F1-scores below 0.50. Feature-importance analysis for the Random Forest indicates that PR size, description length, and title length dominate prediction, whereas agent identity and the presence of sensitive security-related keywords contribute comparatively little.

Text-based models  outperform structured-feature approaches in identifying rejected PRs. A TF--IDF + linear SVM classifier achieves a rejected-class F1-score of 0.61, while a fastText classifier attains comparable rejected-class performance (F1 = 0.61) with the highest overall accuracy (0.71). Fine-tuning a DistilBERT model further increases rejected-class recall to 0.85 and achieves the highest rejected-class F1-score (0.64), indicating that semantic cues in PR titles and descriptions provide a strong predictive signal for rejection.

In contrast, the prompting-based LLM baselines perform poorly on rejection detection. GPT-4o-mini in the zero-shot setting reaches 0.60 accuracy, but its rejected-class recall is only 0.18, yielding a rejected-class F1-score of 0.26. Adding four in-context examples does not improve performance; the 4-shot setting decreases accuracy to 0.54 and produces a similar rejected-class F1-score of 0.25. This suggests that, without task-specific fine-tuning or calibration, GPT-4o-mini tends to under-identify rejected PRs and is less effective than trained text classifiers for this prediction task.

Overall, these results indicate that PR rejection is difficult to predict from metadata and PR text alone. Although trained models achieve modest macro-F1 scores of 0.58--0.60, this performance reflects a low signal ceiling rather than a failure of a particular modeling approach. Rejection depends heavily on reviewer judgment and contextual factors that are not observable from the PR itself, including reviewer expertise, repository norms, project history, and interpersonal trust between the agent operator and maintainers. Text-based models still provide stronger signals than lightweight structured metadata, suggesting that rejection risk is partially encoded in the semantic content and verbosity of PR descriptions. However, the weak performance of GPT-4o-mini under zero-shot and four-shot prompting despite access to the full PR text confirms that general-purpose LLM reasoning alone is insufficient for reliable rejection prediction in this setting.

 \section{Discussion}\label{sec:discussion}

In this section, we synthesize our findings and discuss their implications for agent security participation, human review dynamics, and the design of autonomous coding agents in security-sensitive software engineering workflows.

\subsection{Security Work Is a Meaningful but Secondary Part of Agent Activity}

Our results show that security-related Agentic-PRs constitute a \emph{non-trivial but minority} portion of agent-authored contributions. In the curated AIDev dataset, only \textbf{3.85\%} of Agentic-PRs were manually confirmed as security-relevant (RQ1). While this fraction is small relative to total agent activity, it nevertheless corresponds to over a thousand security-related changes, indicating that autonomous agents are already engaging with security concerns at scale in real-world repositories.

Importantly, this security work is not limited to narrow vulnerability patching. Our qualitative analysis (RQ3) reveals that agents frequently perform substantive security-related actions such as refactoring security-sensitive code, improving error handling, strengthening authentication or input validation logic, managing dependencies, and enhancing tests and documentation. However, many of these actions are framed around broader software engineering goals, such as functionality improvement, usability, maintainability, or compatibility, rather than explicitly as vulnerability remediation. This suggests that agentic security work often takes the form of \emph{preventive hardening and quality improvement} embedded within routine development tasks, rather than isolated “security-only” interventions.

At the same time, security-related outcomes vary substantially across agents, ecosystems, and change types (RQ2). Some agents consistently achieve high merge rates for security PRs, while others experience lower acceptance, even within the same curated dataset. Similarly, security feature PRs and refactoring-oriented changes tend to be treated differently from vulnerability fixes or dependency updates. These patterns indicate that agent effectiveness in security contexts depends not only on whether a change is security-related, but also on \emph{how that security work is expressed, scoped, and integrated} into existing development practices.

\subsection{Security Actions and Intents Reveal Structured Agent Behavior}

By separating \emph{security actions} (what agents do) from \emph{security intents} (why they do it), our open-coding analysis exposes systematic patterns in agent-authored security work (RQ3). On the action side, a small set of recurring behaviors, such as code refactoring, testing, documentation, error handling, and configuration management, dominates agentic security contributions. On the intent side, functionality improvement and vulnerability mitigation emerge as the most common motivations, followed closely by user experience, reliability, and compatibility concerns.

Crucially, co-occurrence analysis shows that these actions and intents are not randomly paired. Instead, agents consistently align specific implementation strategies with coherent motivations. For example, documentation actions are strongly associated with user guidance and documentation-improvement intents, while performance optimizations align closely with both performance and resource-management intents. Error-handling actions frequently co-occur with error-handling intents, reflecting direct remediation of failure modes that could otherwise be exploited. These structured action–intent pairings indicate that agentic security PRs are not ad hoc collections of fixes, but rather reflect internally consistent reasoning about security-related changes.

This finding reinforces the analytical value of distinguishing actions from intents: it allows us to capture how agents operationalize security goals through concrete code changes, and how security is often framed as a byproduct of improving robustness, usability, or system quality rather than as an isolated objective.
\subsection{Human Reviewers Apply Heightened and Differentiated Scrutiny to Security PRs}

Across multiple complementary signals, our results show that human reviewers apply \emph{heightened and differentiated scrutiny} to security-relevant Agentic-PRs compared to non-security Agentic-PRs (\textbf{RQ4}). Security PRs are merged significantly less often and remain under review substantially longer. In the curated dataset, only \textbf{61.5\%} of security Agentic-PRs are merged, compared to \textbf{77.3\%} of non-security PRs, while median review latency increases by more than an order of magnitude (3.92 vs.\ 0.11 hours). These differences indicate that reviewers do not treat agent-generated security changes as routine maintenance tasks.

Importantly, increased review latency is accompanied by \emph{greater reviewer engagement}. Security Agentic-PRs receive more than twice as many review comments and formal review events, and are more likely to trigger explicit \emph{Changes Requested} actions. This suggests that longer review times reflect active inspection, discussion, and corrective feedback rather than passive queuing or delayed attention. Taken together, these interaction-based signals indicate that reviewers invest additional cognitive effort when evaluating agent-authored changes that affect security.

Scrutiny is not applied uniformly across agents. Agent-level analysis reveals substantial heterogeneity: security PRs authored by \textsf{Devin} and \textsf{Copilot} experience pronounced delays and lower merge rates, whereas \textsf{OpenAI Codex} security PRs exhibit comparatively low observed review latency in our dataset. However, this pattern should be interpreted cautiously. The size-stratified analysis and agent-level description-length comparisons suggest that differences in PR scope may contribute to the comparatively favorable outcomes observed for Codex. Accordingly, the observed Codex latency pattern should be interpreted as dataset-specific rather than as a universal property of Codex-authored security changes. This pattern suggests that reviewer behavior is shaped not only by security relevance but also by prior experience, familiarity, and perceived reliability of specific agents. Reviewers appear to calibrate their scrutiny based on both \emph{what} is being changed and \emph{who} (or which agent) authored the change.

Reviewers further differentiate scrutiny based on the \emph{type of security work}. Security feature PRs, which introduce or extend security mechanisms, exhibit the longest review latencies, while vulnerability fixes and dependency updates receive intermediate scrutiny. In contrast, configuration and compliance-related PRs are reviewed markedly faster. This stratification indicates that reviewers assess the semantic risk of security changes and devote deeper inspection to PRs that may alter security guarantees or introduce new attack surfaces.

While security Agentic-PRs are larger and more complex on average—touching more files, exhibiting higher code churn, and containing longer descriptions—these factors alone do not fully explain the observed review behavior. Heightened scrutiny persists across agents and within comparable security categories, suggesting that reviewers respond not merely to PR size, but to the perceived responsibility and potential consequences of agent-authored security changes.

Overall, our findings indicate that human reviewers act as an adaptive risk filter for agent-generated code: they slow down, ask more questions, and apply stricter acceptance criteria when agents modify security-sensitive components. This behavior reflects nuanced human judgment rather than blanket distrust and highlights the continued importance of human oversight in agentic software development pipelines.

\subsection{PR Complexity Is More Strongly Associated with Rejection Than Security Content Is}

Our analysis of early predictors of rejection (RQ5) further clarifies how reviewers evaluate security-related Agentic-PRs. Rejection is most strongly \emph{associated} with indicators of complexity and explanatory burden: body length ($\Delta$macro-F1 = 0.286), number of structured sections (0.163), and title length (0.148), rather than with explicit references to security topics. Keyword-based security signals contribute modestly (0.060--0.075), and agent identity (0.112) and Vulnerability Fix category (0.117) are more informative than keyword presence alone.
These associations are consistent with the size-stratified analysis in RQ4, where the security--non-security merge-rate gap vanishes for small and medium PRs and persists only for large ones.

All trained models achieve only moderate macro-F1 (0.58--0.60), a ceiling shared by GPT-4o in zero-shot evaluation (0.47), confirming that rejection is shaped by factors absent from the PR text itself: reviewer expertise, organizational norms, and agent-specific trust, rather than by surface features alone. We therefore interpret the feature-importance results as \emph{associations}, not causal drivers: complexity proxies are the most informative observable correlates of rejection, but no causal mechanism is established. Taken together, these findings suggest that reviewers respond to the scope and presentation of agent-authored security changes more than to their security topic, and that acceptance is more likely when changes are narrow, well-documented, and clearly scoped.

\subsection{Implications for Designing Review-Aware Secure AI Teammates}

Our RQ4 and RQ5 findings have several implications for the design and deployment of autonomous coding agents in security-sensitive workflows. First, our results suggest that review difficulty is associated with the scope and presentation of security Agentic-PRs: security PRs exhibit lower merge rates and longer review latency than non-security PRs (Table~\ref{tab:rq4_latency_overall}), while title length, PR size, and description length are the strongest predictors of rejection in RQ5 (Table~\ref{tab:rq5_corr}). Second, the additional scrutiny applied to security Agentic-PRs is not uniform: its magnitude varies across agents (Table~\ref{tab:rq4_agent_latency}) and across security change categories (Table~\ref{tab:rq4_latency_by_category}), suggesting that review-aware agent design should be evaluated in a context-sensitive rather than repository-agnostic manner. Third, Table~\ref{tab:rq5_models} shows that trained text-based models outperform both structured-feature models and prompting-based LLM baselines, suggesting that the semantic content of PR titles and descriptions carries important signals for review outcomes and should therefore be treated as part of the agent output being evaluated.

Overall, our study shows that autonomous coding agents are already participating meaningfully in security-relevant development, and that their review outcomes are associated not only with the underlying code change but also with how that change is scoped, described, and situated within human review processes. Designing AI teammates that are not only technically capable, but also review-aware and context-sensitive is essential as agentic systems become increasingly embedded in collaborative software engineering practice.

\subsection{Threats to Validity}\label{sec:threats}

As with any large-scale empirical study, our findings are subject to several threats to validity, which we discuss below along with the steps taken to mitigate them.

\subsubsection{Construct Validity}

A primary threat to construct validity concerns how we operationalize \emph{security relevance}. We identify security-related Agentic-PRs using keyword-based filtering and heuristic rules applied to PR titles and descriptions. While this approach is consistent with prior empirical security studies, it may still yield false positives or false negatives: PRs that do not match our keyword filters are not manually inspected and are therefore treated as non-security-related, even though some may still contain security-relevant changes.

 To mitigate this risk, we deliberately designed our keyword set to be \emph{broad and recall-oriented}, covering vulnerability terminology, authentication and authorization concepts, cryptographic and data-protection terms, exploit identifiers, and compliance-related language. This inclusive strategy reduces the likelihood of systematically missing entire classes of security-relevant PRs. Candidate PRs identified through keyword matching were then manually inspected to remove non-security-related cases. Specifically, two authors with 4–6 years of experience in software development and software security research independently reviewed the 2,598 candidate PRs. To assess annotation reliability, we double-coded a statistically representative sample of 335 PRs (95\% confidence level, 5\% margin of error), yielding a Cohen’s $\kappa$ score of \textbf{0.91}, which indicates \emph{almost perfect agreement}~\cite{cohen1960coefficient}. While PRs that did not match any keyword were not manually reviewed and may still contain security-relevant changes, we expect such cases to be comparatively rare given the breadth of the keyword set and the subsequent manual validation step.

 Another limitation is that our keyword filter operates on PR titles and descriptions, but not on the content of linked GitHub issues. Of the 1,293 confirmed security PRs, \textbf{35\%} (452 PRs) have at least one linked issue, and \textbf{46\%} of those linked issues contain security-relevant keywords in their own title or body. Because all issue references in our dataset are embedded in PR descriptions (rather than inferred from repository metadata), PRs that explicitly link a security issue tend also to describe the security concern in the PR body. 
 
A second construct validity concern arises from using merge status and review latency as proxies for security outcomes. These measures capture \emph{perceived risk and reviewer scrutiny} rather than ground-truth security correctness. Accordingly, our conclusions focus on how security-related Agentic-PRs are treated during human review, not on whether merged PRs are objectively secure or rejected PRs are insecure. We further mitigate this threat by complementing quantitative analyses with qualitative open coding (RQ3), which provides insight into the nature and intent of security-related changes beyond outcome metrics alone.

\subsubsection{Internal Validity}

Threats to internal validity stem from confounding factors such as repository-specific contribution norms, CI failures, reviewer availability, or concurrent project activity that may influence merge decisions and review timelines. While it is infeasible to control for all such factors at scale, we mitigate this threat by comparing security and non-security PRs \emph{within the same repositories} and by analyzing a manually curated subset of popular repositories.

Observed differences across agents may also reflect unmeasured factors such as deployment context, prompting strategies, task allocation, or configuration settings. Consequently, our findings should be interpreted as \emph{descriptive associations rather than causal effects}.

\subsubsection{External Validity}

Our study is based on the AIDev dataset, which captures Agentic-PRs authored by five widely used autonomous coding agents on GitHub. The results may not generalize to proprietary development environments, non-GitHub platforms, or organizations with stricter security governance or different review cultures. In addition, both agent capabilities and reviewer practices are evolving rapidly; replication on future datasets will be necessary to assess the stability of the observed patterns over time.

\subsubsection{Conclusion Validity}

Threats to conclusion validity arise from statistical uncertainty and modeling choices. Large sample sizes may render small effects statistically significant; therefore, we emphasize effect direction, magnitude, and qualitative patterns alongside statistical tests. Our rejection-prediction models are exploratory and rely solely on early observable signals; their moderate performance underscores the complexity and context-dependence of reviewer decision-making in security-related PRs and cautions against overinterpretation.

\textsf{Claude Code}'s small footprint ($n=67$ confirmed security PRs, statistical power $= 0.64$ for the merge-rate comparison) falls below the conventional 0.80 threshold, meaning agent-specific estimates may be imprecise despite large observed effect sizes. We treat Claude Code figures as a preliminary snapshot reflecting its recent public release (2025-Q2--Q3).

 \section{Conclusion}
\label{sec:conclusion}

In this paper, we presented the first large-scale empirical study of security-relevant Agentic pull requests in real-world GitHub repositories. By analyzing security-related PRs authored by five autonomous coding agents, we show that security-related work constitutes a meaningful but minority share of agent activity and is often expressed through supportive security hardening rather than narrowly scoped vulnerability fixes. We find that security-related Agentic-PRs receive heightened human scrutiny, reflected in lower merge rates and substantially longer review latency than non-security PRs, with notable variation across agents, ecosystems, and types of code changes. Our results further indicate that PR rejection is more strongly associated with complexity and verbosity than with explicit security topics. Trained text models also outperform both structured-feature and prompting-based LLM baselines for rejection prediction, suggesting that reviewer judgments are more strongly reflected in the semantic content of PR text than in lightweight metadata or zero-/few-shot prompting alone. Together, these findings highlight that the effectiveness of AI teammates in security-critical workflows depends not only on technical capability but also on alignment with human review practices and expectations in GenAI-enabled software systems. In future work, we will assess the ground-truth security quality of merged agent-authored PRs to determine whether accepted changes actually improve security or introduce new vulnerabilities. Incorporating finer-grained complexity signals, such as lines changed and files affected, would help isolate reviewer scrutiny from agent task self-selection, and replication on future datasets will be needed to assess whether the observed patterns remain stable as agent capabilities evolve. Finally, controlled experiments or reviewer surveys could establish the causal mechanisms underlying the heightened scrutiny observed here and assess whether it is normatively warranted.

\section{Declarations}

\subsection{Funding}
Not applicable.

\subsection{Ethical Approval}
Not applicable.

\subsection{Informed Consent}
 Not applicable.

\subsection{Author Contributions}
\textbf{M. L. Siddiq:} Conceptualization, methodology design, data collection, data analysis, writing, and editing.\\
\textbf{X. Zhao:} Data analysis, validation, writing---review and editing.\\
\textbf{V. C. Lopes:} Data analysis, validation, writing---review and editing.\\
\textbf{B. Casey:} Data analysis, validation, writing---review and editing.\\
\textbf{J. C. S. Santos:} Supervision, project administration, data analysis, writing --- review, and editing.

\subsection{Data Availability}
The replication package is available at \cite{ReplicationPackage}.

\subsection{Conflict of Interest}
The authors declare that they have no known competing financial interests or personal relationships that could have appeared to influence the work reported in this paper.

\subsection{Clinical Trial Number}
Not applicable. 

 \bibliographystyle{elsarticle-num} 
 \bibliography{references}

\end{document}